\documentclass[letterpaper,twocolumn,accepted=2021-11-02]{quantumarticle} 
\pdfoutput=1
\usepackage{amsmath,amsthm,amsfonts,amssymb}

\usepackage{graphicx}
\usepackage{dcolumn}
\usepackage{bm}
\usepackage{hyperref}
\usepackage{amsopn,amsthm,upref,verbatim}
\usepackage{pgfplots}
\pgfplotsset{compat=1.16}
\usepackage[ruled,linesnumbered]{algorithm2e}
\usepackage{mathtools}
\usepackage{textcomp}

\usepackage{float}

\newcommand{\mat}[1]{#1}
\renewcommand{\vec}[1]{#1}
\renewcommand{\mod}{\mathop{\mathrm{mod}}}
\renewcommand{\Pr}{\mathbf{P}}

\newcommand{\abs}[1]{{\lvert#1\rvert}}
\newcommand{\spwt}[1]{{\lvert#1\rvert}_{\mathbf{Sp}}}

\newcommand{\Mat}{\mathcal{M}} 

\newcommand{\F}{\mathbb{F}}
\newcommand{\RC}[2]{\F_{#1}^{\left<#2\right>}} 
\newcommand{\wt}[1]{\mathrm{wt}\left({#1}\right)}
\newcommand{\T}{\mathrm{T}}

\newcommand{\bb}{\mathbbm{b}}

\let\le\leqslant
\let\ge\geqslant

\newcommand{\fI}{\textsc{i}}
\newcommand{\fX}{\textsc{x}}
\newcommand{\fY}{\textsc{y}}
\newcommand{\fZ}{\textsc{z}}

\newcommand{\Iff}{\textbf{if\textcompwordmark f} }
\DeclareMathOperator*{\argmax}{arg\,max}
\DeclareMathOperator*{\argmin}{arg\,min}

\newcommand{\I}{\mathbf{I}}
\newcommand{\C}{\mathbb{C}}

\newcommand{\Z}{\mathbb{Z}}
\newcommand{\N}{\mathbb{N}}
\newcommand{\Stab}{\mathbb{S}}
\newcommand{\zm}{\mathbf{0}}
\newcommand{\cP}{\mathcal{P}}
\newcommand{\cR}{\mathcal{R}}
\newcommand{\cC}{\mathcal{C}}
\newcommand{\cB}{\mathcal{B}}
\newcommand{\cE}{\mathcal{E}}
\newcommand{\cT}{\mathcal{T}}
\newcommand{\cH}{\mathcal{H}}

\newcommand{\cI}{\mathcal{I}}
\DeclareMathOperator{\rk}{rk}

\newenvironment{smallarray}[1]
 {\null\,\vcenter\bgroup\scriptsize
  \arraycolsep=.13885em
  \hbox\bgroup$\array{@{}#1@{}}}
 {\endarray$\egroup\egroup\,\null}

\newcommand{\ket}[1]{\lvert{#1}\rangle}

\DeclareMathOperator*{\Motimes}{\text{\raisebox{0.25ex}{\scalebox{0.8}{$\bigotimes$}}}}

\newlength{\bracewidth}

\theoremstyle{plain}
\newtheorem{lemma}{Lemma}
\newtheorem{proposition}{Proposition}

\theoremstyle{definition}

\theoremstyle{remark}

\newtheorem{example}{Example}
\newtheorem*{example*}{Example}
\newtheorem*{remark*}{Remark}

\makeatletter
\renewcommand*\env@matrix[1][*\c@MaxMatrixCols c]{%
  \hskip -\arraycolsep
  \let\@ifnextchar\new@ifnextchar
  \array{#1}}
\makeatother

\begin{document}

\title{Degenerate Quantum LDPC Codes With Good Finite Length Performance}

\author{Pavel Panteleev}\orcid{0000-0002-6772-7867}
\author{Gleb Kalachev}\orcid{0000-0003-2695-3179}
\affiliation{%
 Faculty of Mechanics and Mathematics, Moscow State University, GSP-1, Leninskie Gory, Moscow, 119991, Russian Federation
}%

\begin{abstract}
We study the performance of medium-length quantum LDPC (QLDPC) codes in the depolarizing channel. Only degenerate codes with the maximal stabilizer weight much smaller than their minimum distance are considered. It is shown that with the help of OSD-like post-processing the performance of the standard belief propagation (BP) decoder on many QLDPC codes can be improved by several orders of magnitude. Using this new BP-OSD decoder we study the performance of several known classes of degenerate QLDPC codes including hypergraph product codes, hyperbicycle codes, homological product codes, and Haah's cubic codes. 
We also construct several interesting examples of short generalized bicycle codes. Some of them have an additional property that their syndromes are protected by small BCH codes, which may be useful for the fault-tolerant syndrome measurement. We also propose a new large family of QLDPC codes that contains the class of hypergraph product codes, where one of the used parity-check matrices is square. It is shown that in some cases such codes have better performance than  hypergraph product codes.
Finally, we demonstrate that the performance of the proposed BP-OSD decoder for some of the constructed codes is better than for a relatively large surface code decoded by a near-optimal decoder. 
\end{abstract}

\maketitle

\section{\label{sc:intro}Introduction}

Quantum error-correcting codes are considered as an~\mbox{essential} component in the current architectures of quantum computers due to the inherently faulty nature of the  quantum hardware. Topological quantum codes~\cite{Kitaev:2002, Freedman:2001, Bombin:2006} are among the quantum codes with the highest known noise thresholds. These codes have sparse parity-check matrices and thus belong to the~class of quantum LDPC (QLDPC) codes. Moreover, they are highly degenerate, which means that the minimum distance is much higher than the weight of the stabilizers. It is important that these codes also have decoding algorithms with near to optimal performance~\cite{Kitaev:2002,Wang:MWM:2011,Bravyi:2014}.   Though the thresholds of topological codes are relatively high, their dimensions are usually much smaller than for general QLDPC codes of the same length (it is constant for the surface and color codes). There are also interesting classes of topological quantum codes with very large dimensions (e.g., hyperbolic codes~\cite{Breuckmann:hyper:2016} have a~constant rate). However, such codes usually have much smaller minimum distances compared to the surface and color codes. 

Recently there have been proposed a number of interesting families of degenerate QLDPC codes (e.g., hypergraph product codes~\cite{Tillich&Zemor:2009} and homological product codes~\cite{Bravyi:HMP:2014}) with very good asymptotic parameters. Nevertheless their practical error-correcting performance for relatively small code lengths ($n < 1000$) is largely unexplored, and it is not clear whether their performance is competitive to the best known topological codes. From our point of view, the difficulty of constructing degenerate QLDPC codes with good practical performance is mostly related to the following two issues. 
\begin{enumerate}
    \item Asymptotically good constructions may not necessarily produce the best QLDPC codes for relatively small code lengths. Indeed, in~\cite{Kovalev&Pryadko:2012, Kovalev&Pryadko:HBP:2013} the construction of hypergraph product codes was further improved and generalized. Although the asymptotic characteristics of the improved codes are the same as before, their parameters such as the~rate and the~minimum distance are much better for smaller lengths. 
    \item The performance of the known QLDPC codes is far from optimal under the state-of-the-art decoders (including the binary and non-binary BP decoders, and their modifications). The performance degradation is usually attributed~\cite{Babar:2015} to the unavoidable $4$-cycles in the corresponding Tanner graphs and to a large number of degenerate errors. While the number of $4$-cycles can be significantly reduced by using CSS codes~\cite{Hagiwara:2007} without $4$-cycles in the~parity-check matrices $H_X$ and $H_Z$, the number of degenerate errors can not be easily reduced for highly-degenerate codes.
\end{enumerate}

In this paper, we try to address both mentioned issues. In the first part of the paper, we introduce a~new enhancement of the standard BP decoder for QLDPC codes (both binary and non-binary versions are allowed) using a~variant of the well-known decoding algorithm for short classical codes called the \emph{ordered statistics decoding (OSD)}~\cite{Fossorier&Lin:1995, Fossorier:2001}. This new post-processing algorithm works only if the BP decoder fails to find a~recovery Pauli operator that gives the correct syndrome. We suppose here that the QLDPC parity-check matrix $H$ is represented in the binary form\footnote{For a QLDPC code on $n$ qubits with $m$ stabilizers the matrix $H$ is an~$m\times 2n$ binary matrix.}. The algorithm starts from finding a~reliable information set~\cite{Dorsch:OSD:1974} for~$H$ based on the soft decisions  obtained by the BP decoder. Then it makes hard decisions for the bits from this information set and flips the $w$ most unreliable of them in order to find the $2^w$ corresponding recovery Pauli operators that return the corrupted quantum state to the coding space. Finally, it selects a recovery operator with the minimum weight and applies it to the corrupted quantum state. Thus this new combined BP-OSD decoder, in contrast to the standard BP, \emph{always} returns the recovery operator that moves the corrupted codeword back to the code space. We show that with the help of this OSD-like post-processing the performance of the standard BP decoder on many degenerate QLDPC codes can be improved by several orders of magnitude. As we can see from its brief description above, it is not restricted to CSS codes and can also be used in conjunction with any decoder, different from BP, that provides soft decisions.  

In the second part of the paper, we construct a~number of new relatively small codes using two families of degenerate codes: the \emph{generalized bicycle} (\emph{GB}) codes introduced in~\cite{Kovalev&Pryadko:HBP:2013} and a~new large family\footnote{Since the first version of the current work was released, the codes from this family have been shown to have very large minimal distances~\cite{Hastings:2021,Panteleev&Kalachev:2020,Breuckmann:balanced:2021} and found interesting applications in geometry~\cite{Freedman:2021}. In~\cite{Panteleev&Kalachev:2020} they were further generalized and called \emph{lifted product} codes.} of QLDPC codes introduced in this paper. We call this new family \emph{generalized hypergraph product} (\emph{GHP}) codes. It contains the GB codes and the class of hypergraph product (HP) codes~\cite{Tillich&Zemor:2009}, where one of the two  parity-check matrices used in the product is square. We also derive new formulas for the dimension of GB and GHP codes and show how to design codes of high dimension with good error correction performance. It is interesting to note that some of the constructed GB codes have an~additional property that their syndromes are protected by small BCH codes, which may be useful for the fault-tolerant syndrome measurement. 

We also study the performance of the proposed BP-OSD decoder on many known classes of degenerate QLDPC codes, including the already mentioned hypergraph product codes, hyperbicycle codes~\cite{Kovalev&Pryadko:2012,Kovalev&Pryadko:HBP:2013}, the homological product codes~\cite{Bravyi:HMP:2014}, and Haah's cubic codes~\cite{Haah:2011}. We compare their performance with the performance of the codes, constructed in this work. We show that in many cases the new codes with similar performance have better parameters such as the code length and the~rate. Besides that, we compare the new BP-OSD decoder with other known modifications of the BP decoder such as the random perturbation~\cite{Poulin:2008}, the enhanced feedback~\cite{Wang:2012}, and the matrix augmentation~\cite{Rigby:2019} algorithms. We compare all the above mentioned algorithms on the new $[[1270, 28]]$ code from the class of GHP codes mentioned earlier\footnote{Since this GHP code is quasi-cyclic, it can also be obtained (up to some qubit permutations) as a~special case of the hyperbicycle code construction.}.  We show that the performance of the new decoding algorithm on this code is significantly better. Moreover, we also show that the performance of this code under the BP-OSD is even better than the performance of the $[[1201,1,25]]$ surface code under the near-optimal MPS-based decoder proposed in~\cite{Bravyi:2014}. We also demonstrate that the performance under BP-OSD of one of~Haah's cubic codes, which have local stabilizers in 3D, is also very good.

The remainder of the paper is organized as follows. Section~\ref{sc:def} contains some background material, where we review standard definitions and fix notations. In~Section~\ref{sc:OSD}, we describe the BP-OSD algorithm and compare its performance with other known modifications of BP. In Section~\ref{sc:gbc}, we study GB codes and construct some new codes with good performance. In Section~\ref{sc:ghp}, we introduce and study a~new large family of GHP codes and show that it generalizes the class of hypergraph product codes in the case when one of the parity-check matrices is square. In the last section, we give some final remarks. The paper also contains three appendixes. Appendix~\ref{app:alg} has some supplementary material on the ring of circulants. Appendix~\ref{app:codes} contains a description of all the codes used in the simulations. Finally, we show some additional simulations in~Appendix~\ref{app:sims}.

\section{\label{sc:def}Basic facts and definitions}

In this section, we fix notations and briefly recall some standard definitions related to classical and quantum LDPC codes. See~\cite{Babar:2015} for a~good review of these topics.

\subsection{Classical codes}

Let $n$ be a natural number. In what follows, we denote by $[n]$ the set $\{1,\dots,n\}$.
Consider a~finite field $\F_q$ and an~\mbox{$n$-dimensional} vector space~$\F_q^n$ over~$\F_q$. A~\emph{linear $[n,k]_q$ code} is a~$k$-dimensional subspace $\cC\subseteq \F_q^n$, where the parameters~$n$ and $k$ are called the~\emph{length} and the~\emph{dimension} of $\cC$, respectively. We denote the~dimension $k$ of the~code~$\cC$ by $\dim \cC$. The~\emph{rate} of the~code~$\cC$ is equal to $k/n$.
The~elements of $\cC$ are called \emph{codewords}. \mbox{The~\emph{Hamming distance} $d(v, v')$} between vectors $v,v'\in \F_q^n$ is the number of positions in which they differ. The parameter 
\[d(\cC) = \min \{d(c, c') \mid c\ne c'; \ c, c'\in \cC\}\] 
is called the \emph{minimal distance} of $\cC$. By definition, we put~$d(\cC)=\infty$ when $k=0$. It is easy to see that $d(\cC)$ is equal to the minimal weight $\abs{c}$ of non-zero codewords, where the~\emph{weight~$\abs{c}$} is the number of non-zero components in~$c$. When $d(\cC)=d$ for a~linear $[n,k]_q$ code $\cC$, we say that $\cC$ is an~$[n,k,d]_q$ code. A~linear $[n, k, d]_q$ code is usually defined either as the row space of a matrix $G$ called the \emph{generator matrix} or as the kernel of a~matrix $H$ called the \emph{parity-check matrix}. It is easy to see that $GH^\T=\zm$,  $\rk G = k$, and $\rk H = n-k$. In what follows we mostly consider \emph{binary} linear codes (i.e., $q=2$), in which case we use a~shorter notation: $[n,k]$ or $[n,k,d]$.

\subsection{Quantum stabilizer codes}\label{sc:stab-codes}

The quantum analogs of classical linear codes are quantum stabilizer codes introduced in~\cite{Gottesman:1997}.  To define them we need a number of supporting definitions. Consider an \emph{$n$-qubit Hilbert space} $\C^{2^n}=(\C^2)^{\otimes n}$. A~\emph{Pauli operator on $n$ qubits} is an operator $P = \alpha P_1\otimes \dots \otimes P_n$ on the space $\C^{2^n}$, where $\alpha\in \{\pm 1, \pm i\}$, $P_i\in\{I,X,Y,Z\}$. Here $I$ is the identity and $X,Y,Z$ are the Pauli $2\times 2$ matrices. The \emph{weight} $\wt{P}$ of a~Pauli operator $P$ is the number of non-identity components in the tensor product. The set $\cP_n$ of all Pauli operators $P$ on $n$ qubits is a non-commutative group under the operator multiplication called the \emph{$n$-qubit Pauli group}. While the group $\cP_n$ is non-commutative, if we forget about the global phase factors $\alpha$ of Pauli operators we obtain the  quotient group $\cP_n/\{\pm \I,\pm i \I\}$, where $\I$ is the identity operator on~$\C^{2^n}$. This quotient group is isomorphic to the commutative group $\Z_2^{2n}$, and the isomorphism is given by the following map:

\begin{equation}\label{eq:stab-mapping}
\alpha \Motimes_{i=1}^n   X^{x_i}Z^{z_i} \mapsto (x_1,\dots,x_n\mid z_1,\dots,z_n).
\end{equation}
It is well known that two $n$-qubit Pauli operators $P$ and $P'$ commute \Iff for their binary representations ${(x|z),(x'|z')\in\F_2^{2n}}$ we have  
\begin{equation}\label{eq:stab-comm}
\left<x,z'\right> +\left<z,x'\right> = 0,
\end{equation}
where $\left<a,b\right>=\sum_i a_ib_i$~is the dot product in~$\F_2^n$.

Denote by $\cP_n^*$ the subset of Pauli operators $P\in \cP_n$ with the global phase factor $\alpha=1$. A \emph{stabilizer group} $\mathcal{S}$ is a~commutative subgroup of the Pauli group $\cP_n$ such that $-\I\not\in \mathcal{S}$. The group~$\mathcal{S}$ is usually generated by $m$ Pauli operators $S_1,\dots,S_m\in\cP^*_n$ called \emph{stabilizers generators}, i.e., $\mathcal{S}=\left<S_1,\dots,S_m\right>$. We say that $S_1,\dots,S_m$ are \emph{independent} if none of them can be obtained (up to a~global phase factor $\alpha$) from the others by the group multiplication.  

Let us recall that a~\emph{quantum stabilizer $[[n, k, d]]$ code} is a~$2^k$-dimensional subspace of the $n$-qubit space $\C^{2^n}$ defined as the common $+1$-eigenspace for a~set of $m$ stabilizers $S_1,\dots,S_m\in \cP_n^*$:
\[ 
    \cC = \{\ket{\psi}\in\C^{2^n} \mid S_i\ket{\psi} = \ket{\psi}, i=1,\dots,m \},
\]
where the group $\Stab(\cC)=\left<S_1,\dots,S_m\right>$ is called the \emph{stabilizer group of $\cC$}. It can be shown that $m \ge n-k$. But if the stabilizers are independent, we get $m=n-k$. Here the parameter $d=d(\cC)$ is called the \emph{minimal distance}\footnote{We write $[[n, k]]$ if the minimal distance is not known.} of $\cC$ and is equal to the minimal possible weight of a~Pauli operator $P\in\cP_n^*$ that commutes with all the stabilizers $S_1,\dots,S_m$ but $P\not\in\Stab(\cC)$. 

A~Pauli operator  $P\in\cP_n^*$ is usually interpreted as an~error operator (called a \emph{Pauli error}) that can corrupt a quantum system and cause it to go from a~state $\ket{\psi}$ to $P\ket{\psi}$. However, for every quantum stabilizer code $\cC$, it is not hard to show that $P\ket{\psi} = \ket{\psi}$ for all $\ket{\psi}\in\cC$ \Iff $P\in \Stab(\cC)$. Hence we see that in this case, not all Pauli errors $P\in\cP_n^*$ can harm the state $\ket{\psi}$. We call a~Pauli error ${P\in\cP_n^*}$ \emph{degenerate} for a~code~$\cC$ if $P\ket{\psi} = \ket{\psi}$ for all $\ket{\psi}\in \cC$ and \emph{non-degenerate} otherwise. We see that the elements from $P\in\Stab(\cC)$ are precisely the degenerate Pauli errors and the minimum distance $d(\cC)$ is the minimal possible weight of a non-degenerate Pauli error.
A~stabilizer code $\cC$ is called \emph{degenerate} if it has degenerate Pauli errors $P$ of weight $\wt{P} < d(\cC)$. 

If we apply the binary mapping (\ref{eq:stab-mapping}) to the stabilizer generators $S_1,\dots,S_m$ of an~$[[n, k, d]]$ code $\cC$, we obtain the~$m\times 2n$ binary matrix 
\begin{equation}\label{eq:stab-parity-check}
H = (H_X \mid H_Z)    
\end{equation}
called the \emph{parity-check matrix} of $\cC$, where each row corresponds to a stabilizer generator. We do not require for the matrix $H$ to be full rank (i.e, $\rk H = n-k \le m$). Using~(\ref{eq:stab-comm}) it is not hard to see that the null space of $H$ is exactly the set of all vectors $(z,x)\in \F_2^{2n}$ such that $(x,z)$ is the~binary mapping of a Pauli error $P\in \cP_n^*$ that commutes with all the~stabilizer generators $S_1,\dots,S_m$.

We see that the matrix $H = (H_X \mid H_Z)$ is not an arbitrary binary $m\times 2n$ matrix since the stabilizer generators $S_1,\dots,S_m$ corresponding to its rows should commute with each other. From equation~(\ref{eq:stab-comm}) it easily follows that this restriction can be formulated as the following \emph{commutativity condition}: 
\begin{equation}\label{eq:comm-cond}
H_X H_Z^\T + H_Z H_X^\T = \zm.
\end{equation}
A~very important subclass of quantum stabilizer codes is \emph{Calderbank-Shor-Steane (CSS) codes} introduced in~\cite{CSS:1996, CSS2:1996}. A~quantum CSS code is a~stabilizer code, where non-identity components in the tensor product of each stabilizer generator are either all equal to $X$ or all equal to $Z$. Hence the parity-check matrix $\mat{H}$ of a~CSS code of length $n$ can be represented in the following form:
\[
\mat{H} =
\begin{pmatrix}[c|c]
\mat{H}_X & \zm\\
\zm & \mat{H}_Z\\
\end{pmatrix},
\]
where $H_X$, $H_Z$ are binary matrices with $n$ columns.

In this special case, commutativity condition~(\ref{eq:comm-cond}) can be rewritten as:
\begin{equation}\label{eq:comm-cond-CSS}
H_X H_Z^\T = \zm.
\end{equation}
We can easily verify that the dimension $k$ of the corresponding CSS code of length $n$ is given by the~formula:
\begin{equation}\label{eq:CSS-dim}
    k = n - \rk H_X - \rk H_Z.
\end{equation}
\subsection{Classical and quantum LDPC codes}

A classical \emph{low density parity check (LDPC) code}~\cite{gallager1963} is a~linear code defined by a~sparse binary parity-check matrix $H=\left(h_{i j}\right)_{m\times n}$. The sparseness usually means that the weights of all rows and columns in $H$ are upper bounded by some universal constant as the code length $n$ grows in an~infinite family of codes. 

When we consider an LDPC code defined by a parity-check matrix $H$, it is helpful to define the bipartite graph $\cT=\cT(H)$ called the \emph{Tanner graph}~\cite{Tanner:1981}. In~this graph the first part of nodes $v_1,\dots,v_n$ (called the \emph{v-nodes}) corresponds to the columns of $H$ (the \emph{variables}), the second part of nodes $c_1,\dots,c_m$ (called the \emph{c-nodes}) corresponds to the rows of $H$ (the \emph{checks}), and we connect a~v-node $v_i$ with a~c-node $c_j$ whenever $h_{i j} = 1$, $i\in[m]$, $j\in [n]$. If the parity-check matrix $H$ is \emph{$(w_c, w_r)$-regular} (i.e., each column has weight $w_c$ and each row has weight $w_r$), then the corresponding Tanner graph is  also \emph{$(w_c, w_r)$-regular} (i.e., each v-node has degree $w_c$ and each c-node has degree $w_r$). We say that an~LDPC code is \emph{$w$-limited} if the degree of each node in its Tanner graph is upper bounded by $w$. It~is obvious that any LDPC code with $(w_c, w_r)$-regular parity-check matrix is $\max(w_c, w_r)$-limited. 

There are a number of decoding algorithms for classical LDPC codes, but the most frequently used one is the \emph{belief propagation (BP) decoder}~\cite{gallager1963}, also known as the \emph{message passing decoder} or the \emph{sum-product decoder}~\cite{Factor_graph:2001}. It assigns the \emph{a~priori}  probability distributions of individual bits in the codeword (obtained from the channel) to the v-nodes of the Tanner graph and iteratively updates the \emph{posterior} probability distributions for each bit. Once some maximal \emph{iteration number limit} is reached, the decoder combines the a~priori and the calculated posterior probability distributions (the~\emph{soft decisions}) to produce the~optimal binary decision (called the \emph{hard decision}) for each individual bit. 

An~important property of a~parity-check matrix $H$ defining an LDPC code is the \emph{girth} of the corresponding Tanner graph $\cT = \cT(H)$, which is equal to the length of the shortest cycle in $\cT$. It is well known that short cycles in the Tanner graph degrade the performance of the BP decoder. At the same time, it was observed that LDPC codes without $4$-cycles (i.e., when the girth is at least $6$) perform very well in practice. However, this practical observation is not fully investigated from theoretical point of view. 

A \emph{quantum LDPC (QLDPC)} is a stabilizer $[[n, k, d]]$ code with a sparse parity-check matrix $H$.
We can also introduce the~Tanner graph $\cT=\cT(\mathcal{S})$ for any stabilizer $[[n, k, d]]$ code $\cC$ defined by a set of stabilizer generators $\mathcal{S}=\{S_1,\dots,S_m\}$. In the case of stabilizer codes, the v-nodes correspond to $n$ qubits and the c-nodes to the stabilizer generators $S_1,\dots,S_m$, and we connect a~c-node with a~v-node if the corresponding stabilizer acts nontrivially on the corresponding qubit.
As in the case of classical LDPC codes, we say that a~QLDPC code is \emph{$w$-limited} if the degree of each node in its Tanner graph is upper bounded by $w$. This property is much more important in the quantum case due to the faulty nature of the current quantum hardware. It is clear that any CSS code with $(w_c, w_r)$-regular matrices $\mat{H}_X$ and $\mat{H}_Z$ is $\max(2w_c, w_r)$-limited.

\section{\label{sc:OSD}OSD-like post-processing for BP}

In this section, we describe  a new OSD-like post-processing algorithm that can be used after the BP decoder for QLDPC codes. Before we give its detailed description we consider two simple modifications of the OSD decoder for \emph{classical} linear codes. These modifications will be used as the main components in the OSD-like post-processing algorithm for quantum codes.
We should warn the reader that these modifications of the standard OSD decoder are not intended to improve its performance for classic LDPC codes. We introduce them because these algorithms eventually will be used in the OSD post-processing algorithm (called qOSD) for \emph{quantum} codes described in Section~\ref{sc:qOSD}. For example, one of the main differences between classical and quantum codes is that we have to use the syndrome decoder in the quantum case. Hence we consider only syndrome OSD decoders here since we are going to use them as components of the decoder for quantum codes.

\subsection{Syndrome OSD post-processing algorithm}\label{sc:osd-w}

Starting from this section we will often use the following notations:
\begin{itemize}
    \item If $M$ is a~matrix, then $M_i$ denotes its~$i$-th column.
    \item If $I=\{i_1,\dots,i_k\}$ is some~index set, $i_1 < \dots < i_k$,  and $\pi\in\mathbf{S}_n$ is a permutation, then for every vector $v=(v_1,\dots,v_n)$ and matrix $M=(M_1,\dots,M_n)$ we define: 
    \begin{align*}
    v_I &= (v_{i_1},\dots,v_{i_k}),\\
    M_I &= (M_{i_1},\dots,M_{i_k}),\\
    \pi(v) &= (v_{\pi(1)},\dots,v_{\pi(n)}),\\ 
    \pi(M) &= (M_{\pi(1)},\dots,M_{\pi(n)}).    
    \end{align*}
    \item If $I\subseteq [n]$ is an~index set, then the index set $\bar{I} = [n]\setminus I$ is called its \emph{complement}.
\end{itemize}

Now let us recall the definition of an~information set~\cite{Prange:1962} of a~code, which has an~important role in the~OSD decoding. A~set of indices $I\subseteq[n]$ is called an~\emph{information set} of a~classical linear $[n,k]$ code~$\cC$ if $\cC_I = \{c_I\mid c\in\cC\} = \F_2^k$. Clearly, $I$ is an~information set of $\cC$ \Iff $\abs{I}=k$, and for any two codewords $c,c' \in \cC$ such that  $c_I = c'_I$ we always have $c = c'$. Therefore for any information set~$I$ the corresponding $k$ bits can be used to \emph{uniquely} recover any vector $v\in\F_2^n$ if we also know its syndrome 
$s=Hv$.   Hence we can consider the corresponding \emph{encoding map} $\cE^s_I\colon \F_2^k \to \F_2^n$ such that for any $u$, $v$, and $s$ we have:
\[
v_I=u,\ Hv = s \quad\Longleftrightarrow\quad v=\cE^s_I(u).
\]
It is easy to verify that $\cE^\zm_I$ is the~systematic encoding map for the~code~$\cC$ where the information bits $u$ are at the~positions corresponding to $I$. 
\begin{remark*}
If $G$ is a~generator matrix, and $H$ is a~parity-check matrix of a~linear code~$\cC$; then it can be shown that a~$k$-element index set $I$ is an~information set of~$\cC$ \Iff $\rk G_I = \rk G$ \Iff $\rk H_{J} = \rk H$, where $J = \bar{I}$. Thus if $\cB(M)$ denotes the family of indices   $I$ such that the~collection of columns $\{M_i\}_{i\in I}$ is a~basis for the column space of~$M$; then the~family of all information sets of~$\cC$ coincides with $\cB(G)$, while the family  of their complements coincides with $\cB(H)$. Since the~set of all linearly independent columns of any matrix gives us a~matroid\footnote{A~\emph{matroid} is defined by a~non-empty collection $\cI$ of subsets (called the \emph{independent sets}) from some set $E$ (called the \emph{ground set}) such that: (1) $\cI$ is closed under taking subsets; (2) for any two $A, B\in \cI$ if $\abs{A} < \abs{B}$ then $A\cup\{b\}\in \cI$ for some $b\in B\setminus A$. Any~maximal (by inclusion) independent set is called a~\emph{basis}.}, then for any positive real numbers $w_1,\dots,w_k$ we can \mbox{efficiently} find 
\begin{equation}\label{eq:matroid}
\argmax_{I\in\cB(G)} \sum_{i\in I} w_i = \argmin_{J\in\cB(H)} \sum_{i\in J} w_i
\end{equation}
in a~greedy fashion~\cite{Edmonds:1971,Fossorier:1998}. Here in the right part we used that $I\in\cB(G)$ \Iff $J = \bar{I} \in \cB(H)$.   
\end{remark*}
The main idea of the~syndrome OSD decoder is as follows. Consider a~linear $[n,k]$ code $\cC$ defined by a~parity-check matrix $H$. Let $c'= c + e$ be a~corrupted version of a~codeword ${c\in\cC}$, where ${e\in\F_2^n}$ is the corresponding random \emph{error vector}. Given the syndrome $s = H c' = H e$ we want to find the error vector $e$ and thus recover the codeword $c = c' - e$. Suppose that in addition we are given an estimate\footnote{For example, one can use a~syndrome BP decoder for this purpose (see Section~\ref{sc:BP-OSD}).} of the error probability $p_i = \Pr(e_i = 1)$ for each $i\in[n]$. From these estimates our best guess of the~error vector $e$ would be the \emph{hard decisions} vector~$\hat{e}$, where  $\hat{e}_i = 1$ when $p_i > 1/2$, and $\hat{e}_i = 0$ otherwise\footnote{When $p_i=1/2$ we can set $\hat{e}_i$ randomly to either $0$ or $1$ with equal probability.}. However, when we only use the error probability estimates, we often have that $H\hat{e} \ne s$, and thus $\hat{e} \ne e$. Nevertheless, even in such cases, some components of $\hat{e}$ are equal to~$e$, and we can still try to use $\hat{e}$ to find $e$ if we also take into account that
\begin{equation}\label{eq:syndrome}
He = s.      
\end{equation}
Therefore, to recover $e$, one may try to traverse through different information sets $I$ in the hope that for one of them we have $\hat{e}_I = e_I$, and thus 
\begin{equation}\label{eq:osd-sol}
e = \cE_I^s(\hat{e}_I).     
\end{equation}
Unfortunately we do not know for which indices $i\in[n]$ we have $\hat{e}_i = e_i$. Hence it makes sense to find an~information set $I$ with the indices that are as reliable as possible, where the~\emph{reliability} of an~index $i\in[n]$ is the~probability 
\[
\rho_i = \Pr(\hat{e}_i = e_i) = \max(p_i, 1 - p_i).
\]      
If we assume that the components in the random error vector $e$ are mutually independent, then it follows that the probability of successful decoding  $\Pr(e = \cE_I^s(\hat{e}_I))$ for $I$ is equal to $\rho(I)=\prod_{i\in I} \rho_i$. Thus if we want to maximize this probability we need the \emph{most reliable information set~$I$}, i.e., the one\footnote{If there are several such sets, we can use any of them.} with the maximal possible value of $\rho(I)$. Finding such a~set may at first sight look like a~prohibitively hard task. However, if instead of $\rho(I)$ we consider $\ln \rho(I) = \sum_{i\in I} w_i$, where $w_i = \ln \rho_i$; then we can see that the most reliable information set~$I$ is given by~(\ref{eq:matroid}), and, as we mentioned earlier, it is possible to find $I$ using a~greedy algorithm~\cite{Edmonds:1971,Fossorier:1998}.

This greedy algorithm is the main part of the OSD decoder. Since in our case the code is defined by a~parity-check matrix $H$, it is more convenient to use the right part of (\ref{eq:matroid}), which means that we want to find the index set $J$ corresponding to the \emph{least reliable basis} $\{H_i\}_{i\in J}$ for the column space of~$H$, i.e., $J\in\cB(H)$ with the smallest~$\rho(J)$. For simplicity, we assume that \emph{the columns of $H$ are already rearranged} such that the reliability of the corresponding positions increases: $\rho_1 \le \dots \le \rho_n$. In~this case, it is not hard to see that the collection $\{H_i\}_{i\in J}$ of the first $r= \rk H$ linearly independent columns is the least reliable basis~\cite{Fossorier:1998}. We can find $J$ by applying Gaussian elimination to equation (\ref{eq:syndrome}), which gives us, in time $O(n^3)$, the following equation: 
\begin{equation}\label{eq:gauss-syndrome}
\tilde{H} e = \tilde{s}.    
\end{equation}
Then $J$ is the index set of the first $r$ pivot columns in $\tilde{H}$, i.e., $\tilde{H}_J = (\delta_{i j})_{r\times r}$. Moreover, from~(\ref{eq:matroid}) it follows that the index set~$I = \bar{J}$ is the most reliable information set, and we can find the error vector $e$ from (\ref{eq:gauss-syndrome}) in a~very straightforward way since $\tilde{H}_J$ is the identity matrix.

\begin{algorithm}[t]
    \KwIn{target weight function $\wt{\cdot}$,\\binary parity-check matrix $H$,\\ syndrome vector $s\in \F_2^m$,\\ vector of hard decisions  $\hat{e}\in \F_2^n$;}
    \KwOut{error vector $e\in \F_2^n$ such that $H e = s$;}
    $J\leftarrow \varnothing$\label{walg:gauss-b}\;
    \For{$i\leftarrow 1$ \KwTo $n$}{
        $J'\leftarrow J\cup\{i\}$\;
        \If{$\rk H_{J'}>\rk H_J$}{
            $J\leftarrow J'$\;
        }
    }\label{walg:gauss-e}
    $\hat{x} \leftarrow\argmin\limits_{x\in\F_2^w}\mathrm{wt}\bigl(\cE_I^s(\cR_{[w]}^x\hat{e}_I)\bigr)$, where $I = \bar{J}$\label{walg:osd-list}\;
    \Return $e = \cE_I^s(\cR_{[w]}^{\hat{x}}\hat{e}_I)$\;
    \caption{Syndrome OSD-$w$ algorithm}\label{alg:osd-w}
\end{algorithm}

The described above syndrome decoding algorithm is usually called \emph{order-$0$ OSD decoder}. As we see, it first finds the most reliable information set $I$ and then recovers the unique vector $e\in\F_2^n$ subject to the following conditions: 
\begin{equation}\label{eq:osd-0}
e_I = \hat{e}_I,\quad He = s.     
\end{equation}
In~fact, we can further improve the error-correcting performance by using \emph{order-$w$ OSD decoder} (abbreviated as OSD-$w$). In this modification, after we find the most reliable information set~$I$, we look at all error vectors $e$ obtained from equation~(\ref{eq:gauss-syndrome}) by setting the first $w$ least reliable positions from $I$ to all possible values $x\in\F_2^w$ (we can obtain them using the encoding operator $\cE_I^s$). Finally, we select an~error vector $e$ of the~minimal weight $\wt{e}$ among all the~$2^w$ obtained error vectors. Here the~weight function $\wt{\cdot}$ depends on the~specific channel we use. In the case of depolarizing noise, this weight function is just the weight of the corresponding Pauli operator.

\begin{algorithm}[t]
    \SetKw{Break}{break}
    \KwIn{binary parity-check matrix $H$,\\ syndrome vector $s\in \F_2^m$,\\ vector of hard decisions $\hat{e}\in \F_2^n$;
    }
    \KwOut{error vector $e\in \F_2^n$ such that $H e = s$;}
    $J\leftarrow \varnothing$\;
    $s'\leftarrow s+H \hat{e}$\;
    \For{$i\leftarrow 1$ \KwTo $n$}{
            \If{$\rk {[H_J,s']}=\rk H_J$}{
                \Break \;
            }
            $J'\leftarrow J\cup\{i\}$\;
            \If{$\rk H_{J'}>\rk H_J$}{
            $J\leftarrow J'$\;
            $s'\leftarrow s'+\hat{e}_{i}H_{i}$\;
        }
    }
    $x\leftarrow$ the~solution of $H_J\, x=s'$\;
    \Return $e = \cR_J^x \hat{e}$\;
    \caption{Fast syndrome OSD-$0$ algorithm}\label{alg:osd0}
\end{algorithm}

A~simplified pseudocode of the above OSD-$w$ algorithm is shown in~Algorithm~\ref{alg:osd-w}. Here in the two last lines, we used a~shorthand notation $\cR_I^x v$ for the result of the replacement of the~subvector $v_I$ in the vector $v$  by $x\in\F_2^{\abs{I}}$. Therefore, $\cR_{[w]}^x\hat{e}_I$ is obtained from $\hat{e}_I$ if we replace its first $w$ bits by $x\in \F_2^w$. Since we assume that the columns of $H$ are already sorted according to the reliabilities, these first $w$ positions in $I$ are the least reliable ones.  We should also note that the main cycle of this algorithm, which finds $J$ (lines~\ref{walg:gauss-b}--\ref{walg:gauss-e}), can be efficiently implemented using Gaussian elimination, as already discussed above.

\begin{remark*}
There are some differences between the standard order-$w$ OSD decoder from~\cite{Fossorier&Lin:1995, Fossorier:2001} and the~proposed OSD-$w$ post-processing algorithm. For example, we try all $2^w$ bit flips of the $w$ least reliable information bits (line~\ref{walg:osd-list}), while in the standard OSD we try all the $\sum_{i\le w}\binom{k}{i}$ bit flips of no more than $w$ bits. 
\end{remark*}

As we will see in~Section~\ref{sc:BP-OSD}, if the OSD-$w$ decoder is used as a~post-processing algorithm after the BP decoder for QLDPC codes, then it can radically improve the error correcting performance. In fact, in many cases even OSD-$0$ is enough, and thus the computational cost of the decoding is $O(n^3)$. Moreover, in the case of the \mbox{OSD-$0$}, there is no need to do full Gaussian elimination for $H$. Let $I$ be the most reliable information set found by the OSD-$0$ algorithm, then we can stop extending the index set~$J$ in~Algorithm~\ref{alg:osd0} when $\rk {[H_J,s']}=\rk H_J$, and put $e = \cR_J^x \hat{e}$, where $s' = s + H_{\bar{J}}\hat{e}_{\bar{J}}$, and $x$ is the~unique\footnote{The solution is unique since $\rk H_J = \abs{J}$.} solution of $H_J x = s'$. Indeed, we get
\[
He = H_J x + H_{\bar{J}} \hat{e}_{\bar{J}} = s' + H_{\bar{J}} \hat{e}_{\bar{J}} = s. 
\]
Since we also have $I\subseteq \bar{J}$, then $e$ satisfies conditions~(\ref{eq:osd-0}) and thus is the error vector found by the OSD-$0$ algorithm. This observation gives us a~faster version of the OSD-0 algorithm (see Algorithm~\ref{alg:osd0}).

\subsection{Modified OSD post-processing algorithm for stabilizer codes}\label{sc:qOSD}

In this subsection, we show how to adapt the OSD decoder from the previous subsection to a~scenario where it is used as a~post-processing algorithm after the BP decoder for QLDPC codes. 
There are a number of quantum noise models. In this paper, we consider the depolarizing channel only. However many of our ideas may be used for other memoryless quantum noise models. In the \emph{depolarizing channel} model with \emph{error probability~$p$}, a~quantum state ${\ket{\psi}\in \C^{2^n}}$ is subject to a~random Pauli error 
\[
E=E_1\otimes\dots\otimes E_n\in \cP_n^*,
\]
where  all $E_i$ are i.i.d, and for all $i\in [n]$ we have: 
\[\Pr(E_i = X) = \Pr(E_i = Y) =  \Pr(E_i = Z) = p/3.\]

In what follows, it is convenient for our goals to represent (with a~small abuse of notation and terminology) the set of matrices $\cP_1^*=\{I,X,Y,Z\}$ by the elements of the finite field~$\F_4$, where $I$ is represented by $0\in\F_4$, and the Pauli matrices $X,Y,Z$ by the three non-zero elements from~$\F_4$. To distinguish these finite field elements from the corresponding matrices we denote the former using a~different font: $\fI, \fX, \fY, \fZ$. Further, we represent a~Pauli vector from $\cP_n^*$ by the corresponding vector from~$\F_4^n$, which we also call a~\emph{Pauli vector}. Using this conventions, we represent the binary $m\times{2n}$ parity-check matrix of a~stabilizer code $\cC$ (see equation~(\ref{eq:stab-parity-check})) by the corresponding $m\times n$ matrix over $\F_4$ and call it the \emph{stabilizer matrix} of $\cC$.

Since we consider the depolarizing channel, for the best performance, it is better to use the non-binary version of the syndrome BP decoder~(see~\cite{Poulin:2008}, \cite[Algorithm 1]{Babar:2015}), which also takes into account the correlations between $X$ and $Z$ errors in qubits.
To describe the OSD algorithm in this case we need some extra notations.
\begin{itemize}
    \item If $v\in \F_4^n$ is a~Pauli vector, and $v_i=v^X_i \fX  + v^Z_i \fZ$, where $v^X_i,v^Z_i\in \F_2$, $i\in[n]$; then we can define the~binary vectors:  
    \begin{align*}
    \bb(v) & = (v^X_1, v^Z_1,v^X_2,v^Z_2,...,v^X_n,v^Z_n)\in \F_2^{2n},\\ 
    \bb^*(v) & = (v^Z_1, v^X_1,v^Z_2,v^X_2,...,v^Z_n,v^X_n)\in \F_2^{2n}.      
    \end{align*}
    We also need the inverse mapping $\bb^{-1}(\cdot)$ for $\bb(\cdot)$ that maps vectors from $\F_2^{2n}$ back to $\F_4^n$.
    \item If $\cH$ is an~$m\times n$ stabilizer matrix over $\F_4$, then $\bb(\cH)$ denote the~$m\times 2n$ binary matrix obtained by mapping each row $h$ from $\cH$ to the row $\bb^*(h)$.  
\end{itemize}

The main motivation for these, not very standard, notations is as follows.
If ${e\in\F_4^n}$ is a~Pauli error, then it is easy to check that
\[s=\bb(\cH)\bb(v)\]
is the corresponding \emph{syndrome} vector, i.e., for every ${i\in [m]}$ we get  $s_i=0$ \Iff the~$i$-th stabilizer from $\cH$ commutes with the Pauli error~$e$. Let us emphasize that the binary representations $\bb(v)$ and $\bb(\cH)$ can be also obtained from the binary representations (\ref{eq:stab-mapping}) and (\ref{eq:stab-parity-check}) by a~permutation of indices and columns, respectively.

\begin{example*}
Let us illustrate the above notations on the well-known non-CSS $[[5,1,3]]$ code~\cite{Laflamme:1996} defined by the parity-check matrix
\[
H = (H_X |H_Z) = 
\left(\begin{smallarray}{ccccc|ccccc}
   1 & 0 & 0 & 1 & 0\hphantom{|} & \hphantom{|} 0 & 1 & 1 & 0 & 0\\
   0 & 1 & 0 & 0 & 1\hphantom{|} & \hphantom{|} 0 & 0 & 1 & 1 & 0\\
   1 & 0 & 1 & 0 & 0\hphantom{|} & \hphantom{|} 0 & 0 & 0 & 1 & 1\\
   0 & 1 & 0 & 1 & 0\hphantom{|} & \hphantom{|} 1 & 0 & 0 & 0 & 1
\end{smallarray}\right),
\]
which corresponds to the matrices
\[
	\cH = \left(\begin{smallmatrix}
	\fX & \fZ & \fZ & \fX & \fI\\
	\fI & \fX & \fZ & \fZ & \fX\\
	\fX & \fI & \fX & \fZ & \fZ\\
	\fZ & \fX & \fI & \fX & \fZ\\
	\end{smallmatrix}\right),
	\bb(\cH) = \left(\begin{smallmatrix}
	0 & 1\hphantom{|} & 1 & 0\hphantom{|} & 1 & 0\hphantom{|} & 0 & 1\hphantom{|} & 0 & 0\\
	0 & 0\hphantom{|} & 0 & 1\hphantom{|} & 1 & 0\hphantom{|} & 1 & 0\hphantom{|} & 0 & 1\\
	0 & 1\hphantom{|} & 0 & 0\hphantom{|} & 0 & 1\hphantom{|} & 1 & 0\hphantom{|} & 1 & 0\\
	1 & 0\hphantom{|} & 0 & 1\hphantom{|} & 0 & 0\hphantom{|} & 0 & 1\hphantom{|} & 1 & 0\\
	\end{smallmatrix}\right).
\]
If we have the $Z$ error in the second qubit, then we get the error vector $e=(\fI, \fZ, \fI, \fI, \fI)\in\F_4^5$, its binary representation $\bb(e)=(0,0,0,1,0,0,0,0,0,0)\in\F_2^{10}$, and the corresponding syndrome vector
$s = (0,1,0,1)$.

\end{example*}

Since the depolarizing channel is non-binary, we need some further adjustments in the syndrome OSD decoder (Algorithms~\ref{alg:osd-w} and \ref{alg:osd0}) to use it as a~post-processor after the non-binary BP decoder. We call this modified version the \emph{order-$w$ qOSD} algorithm, which is defined via the order-$w$ OSD algorithm from the previous section as follows:
\[
\mathrm{qOSD}_w(\cH,s,\hat{e})=\bb^{-1}\bigl(\mathrm{OSD}_w(\spwt{\cdot},\bb(\cH),s,\bb(\hat{e}))\bigr).
\]
Here we use the \emph{symplectic weight}\footnote{It is easy to see that the symplectic weight $\spwt{x}$ is equal to the weight $\wt{E}$ of the Pauli error $E\in\cP_n^*$ represented in the binary form by the vector $x\in\F_2^{2n}$.}  $\spwt{\cdot}$ as the target weight function $\wt{\cdot}$ for the OSD algorithm, which is defined as:
\[
\spwt{x}=\sum_{i=1}^n (x_{2i-1}\vee x_{2i}).
\]
As we see, the input for the qOSD algorithm includes the $m\times n$ stabilizer matrix $\cH$ over $\F_4$, the syndrome $s\in \F_2^m$, and the error vector $\hat{e}\in\F_4$, which is the hard decisions vector for the result of the non-binary BP decoder (see~the next subsection).

\begin{remark*}
The symplectic weight $\spwt{\cdot}$ is used above for the depolarizing channel. For the~variant of the depolarizing channel, where the $X$ and $Z$ errors are independent, the standard binary Hamming weight $\abs{\cdot}$ should be used instead. In general, the optimal choice of the target weight function $\wt{\cdot}$ is the one, where the most probable errors are of the least possible weight.  
\end{remark*}

	
	
\subsection{\label{sc:BP-OSD}Main decoding algorithm (BP-OSD)}

In this subsection, we describe our main decoding algorithm (see Algorithm~\ref{alg:osd-main}) called the \emph{BP-OSD} or \emph{BP-OSD-$w$} (if we want to emphasize the OSD order $w$). It consists of two stages: the~non-binary BP decoding (lines~\ref{walg:soft}--\ref{walg:syndrome}) and the~OSD post-processing (lines~\ref{walg:bs}--\ref{walg:es}). In~fact, at the first stage, any decoding algorithm may be used instead of BP if it can provide soft decisions. The goal of this stage is either to find the error vector $e\in\F_4^n$ or to obtain the~probabilities $\Pr(e_i = E)$, $i\in [n]$, for each type of the~Pauli error $E\in \{\fI,\fX,\fY,\fZ\}$. 

First, we run the BP decoder and obtain the \emph{soft decisions} (line~\ref{walg:soft}), i.e., for ever $i\in[n]$ we get the $4$ numbers $(p_{i,\fI},p_{i,\fX},p_{i,\fY},p_{i,\fZ})$, where $p_{i,E}$ can be interpreted as the~probability $\Pr(e_i=E)$ of the error $E\in \{\fI,\fX,\fY,\fZ\}$ in the~$i$-th qubit. Next, we find the \emph{hard decisions} for all qubits by the~formula (line~\ref{walg:hard}):
\[
\hat{e}_i=\argmax_{E\in\{\fI,\fX,\fY,\fZ\}}p_{i,E}, i\in[n].
\]
If the~BP decoding is successful (i.e., we have the correct syndrome vector), then there is no need in the~post-processing, and the~BP-OSD decoder returns the~result of the BP decoder (line~\ref{walg:syndrome}). 

After the first stage, the probabilities $p_{i,E}$ are also used in the BP-OSD algorithm to sort the qubits in the~order of increasing reliability $\rho(p_{i,\fI},p_{i,\fX},p_{i,\fY},p_{i,\fZ})$, where we can define the \emph{reliability} of the $i$-th qubit as 
\[
\rho(p_{i,\fI},p_{i,\fX},p_{i,\fY},p_{i,\fZ}) = \Pr(\hat{e}_i = e_i) =\! \max_{E\in\{\fI,\fX,\fY,\fZ\}}{p_{i,E}}.
\]
However, in all our simulations we used the formula 
\[
\rho(p_{i,\fI},p_{i,\fX},p_{i,\fY},p_{i,\fZ}) = p_{i,\fI},
\]
which gives similar error-correcting performance, but it is a~little bit easier to calculate. After we sorted the qubit positions (line~\ref{walg:sort}), the~qOSD algorithm from the previous subsection is used to recover the errors in the~least reliable qubits from the~hard decisions for the~remaining qubits (line~\ref{walg:es}).

\begin{algorithm}
	\SetKwInOut{Input}{input}\SetKwInOut{Output}{output}\SetKwInOut{Parameter}{Parameters}
	\SetAlgoLined
	\KwIn{stabilizer matrix $\cH$,\\ syndrome vector $s\in \F_2^m$;}
	\KwOut{error vector $e\in \F_4^n$;}
	$(p_{i,\fI},p_{i,\fX},p_{i,\fY},p_{i,\fZ})_{i=1}^n\leftarrow \mathrm{BP}(s)$\label{walg:soft}\;
	Make hard decisions:\quad
	$\hat{e}_i \leftarrow \argmax_{E\in\{\fI,\fX,\fY,\fZ\}} p_{i,E}$, $i\in[n]$\label{walg:hard}\;
	\eIf{$\bb(\cH)\bb(\hat{e}) = s$}{\Return $e=\hat{e}$\;\label{walg:syndrome}}{
		Calculate the reliabilities: $\rho_i\leftarrow \rho(p_{i,\fI},p_{i,\fX},p_{i,\fY},p_{i,\fZ})$\label{walg:bs}, $i\in[n]$\;
		Sort the qubits by their reliabilities:
		$\rho_{\sigma(1)} \le \rho_{\sigma(2)} \le ...\le \rho_{\sigma(n)}$, $\sigma\in S_n$\label{walg:sort}\;
		\Return $e=\sigma^{-1}\bigl(\mathrm{qOSD}_w (\sigma(\cH),s,\sigma(\hat{e}))\bigr)$\label{walg:es}\;
	}
	\caption{BP with OSD-$w$ post-processing}\label{alg:osd-main}
\end{algorithm}

In Algorithm~\ref{alg:osd-main}, we used the qOSD algorithm, which is a~good choice for the standard depolarizing channel. If we have a~channel with independent $X$ and $Z$ errors, then for a~CSS code defined by a~pair of parity-check matrices $H_X,H_Z$ we can also use the~standard OSD (Algorithms~\ref{alg:osd-w} and~\ref{alg:osd0}) for the~$X$ and $Z$ components of the error vector $e$ separately. In this case, lines \ref{walg:bs}--\ref{walg:es} in Algorithm~\ref{alg:osd-main} can be replaced by the~following steps:
\begin{enumerate}
    \item Calculate the $X$ and $Z$ error probabilities:
    \begin{align*}
        p_i^X&=p_{i,\fX}+p_{i,\fY},\\
        p_i^Z&=p_{i,\fY}+p_{i,\fZ}.
    \end{align*}
    \item Sort the qubits in the \emph{decreasing} order of their $X$ error and $Z$ error probabilities and let $\sigma_X,\sigma_Z\in S_n$ be the corresponding permutations such that: 
    \begin{align*}
    &p_{\sigma_X (1)}^X\ge p_{\sigma_X (2)}^X\ge ...\ge p_{\sigma_X (n)}^X,\\
    &p_{\sigma_Z (1)}^Z\ge p_{\sigma_Z (2)}^Z\ge ...\ge p_{\sigma_Z (n)}^Z.
    \end{align*}
    \item Let $\hat{e}=\hat{e}_X \fX + \hat{e}_Z \fZ$, $s_X = H_X \hat{e}_Z$, and $s_Z = H_Z \hat{e}_X$. It is clear that $s = [s_X,s_Z]$. Run the~OSD decoder separately for the $X$ and $Z$ components of $\hat{e}$:
    \begin{align*}
        e_X&=\sigma_X^{-1}\bigl(\mathrm{OSD}_w (|\cdot|,\sigma_X(H_Z),s_Z,\sigma_X(\hat{e}_X))\bigr),\\
        e_Z&=\sigma_Z^{-1}\bigl(\mathrm{OSD}_w (|\cdot|,\sigma_Z(H_X),s_X,\sigma_Z(\hat{e}_Z))\bigr).
    \end{align*}
    \item Return $e=e_X \fX+e_Z \fZ$.
\end{enumerate}

For all our simulations we used the normalized min-sum (NMS) decoder with the~normalization factor $0.625$, which approximates the non-binary BP decoder from~\cite{Poulin:2008}, \cite[Algorithm 1]{Babar:2015} in log-domain and is more numerically stable in some cases. The maximal number of iterations was set to $32$. We used the layered scheduling in order to increase the convergence speed of the decoder by approximately two times\footnote{Since the first version of the current work was released, the layered (serial) schedule of the BP decoder for QLDPC codes was also studied in~\cite{Kuo:2020}.}. For a~good review of practical aspects of the BP decoder implementation, see~\cite[Chapter~4]{ldpc:book:2014}. The error-correcting performance in our simulations is measured either in terms of the \emph{logical error rate} or the \emph{word \mbox{error} rate}\footnote{Recall that the \emph{word error rate}, also known as the \emph{block error rate} or \emph{frame error rate}, is the rate of codewords where the decoder does not give a~correct answer (i.e., it either fails to decode or we have at least one logical error after the decoding).} (\emph{WER}).  We should also stress that in many cases we  use only the OSD-0 algorithm (Algorithm~\ref{alg:osd0}), which has complexity $O(n^3)$, while the complexity in the general case is $O(n^3+n2^w)$.

\begin{figure}
\begin{tikzpicture}
\pgfplotsset{major grid style={dashed}} 
\pgfplotsset{minor grid style={dotted}}
\pgfplotsset{
    tick label style={font=\footnotesize},
    label style={font=\footnotesize},
    legend style={font=\scriptsize},
    standard/.style={
        every axis y label/.style={at={(ticklabel cs:0.5)}, rotate=90,anchor=center,font=\footnotesize},
        every axis x label/.style={at={(ticklabel cs:0.5)}, anchor=center,font=\footnotesize}
    }
}
\begin{loglogaxis}[
ylabel=WER,
xlabel=Physical error rate,
ymax=1,
ymin=1e-8,
xmin=0.01,
xmax=0.15,
width=\linewidth,
legend columns=2,
legend style={
	cells={anchor=west},
	legend pos= north west,
},
every axis y label/.style=
{at={(ticklabel cs:0.5)},rotate=90,anchor=center,font=\footnotesize},
every axis x label/.style=
{at={(ticklabel cs:0.5)},anchor=center,font=\footnotesize},
grid=both,
title=The effect of OSD-0 for different QLDPC codes,
enlarge y limits=.01,
height=8cm,
]%
\addplot[black,mark=o] table[x=P,y=WER] {sims/res_gbc55-254_28_14.txt};
\addlegendentry{\ref{code:gbc-254_28}, BP}
\addplot[black,mark=*] table[x=P,y=WER] {sims/res_gbc55-254_28_14-osd.txt};
\addlegendentry{\ref{code:gbc-254_28}, BP-OSD}
\addplot[red,mark=o] table[x=P,y=WER] {sims/res_ghpc33-882_24-noosd.txt};
\addlegendentry{\ref{code:ghpc33-882_24}, BP}
\addplot[red,mark=*] table[x=P,y=WER] {sims/res_ghpc33-882_24-osd0.txt};
\addlegendentry{\ref{code:ghpc33-882_24}, BP-OSD}

\addplot[green!70!black,mark=o] table[x=P,y=WER] {sims/res_hpc55long-noosd.txt};
\addlegendentry{\ref{code:hpc55-7938_578_16}, BP}
\addplot[green!70!black,mark=*] table[x=P,y=WER] {sims/res_hpc55long-osd.txt};
\addlegendentry{\ref{code:hpc55-7938_578_16}, BP-OSD}

\addplot[blue,mark=o] table[x=P,y=WER] {sims/res_haah-1024_30-nosd.txt};
\addlegendentry{\ref{code:haah8}, BP}
\addplot[blue,solid,mark=*] table[x=P,y=WER] {sims/res_haah-1024_30.txt};
\addlegendentry{\ref{code:haah8}, BP-OSD}

\end{loglogaxis}
\end{tikzpicture}

\caption{The WER performance of several QLDPC codes under the BP and BP-OSD-0 decoders, where: \ref{code:gbc-254_28} is the~$10$-limited generalized bicycle [[254,28]] code;
\ref{code:ghpc33-882_24} is the~$6$-limited generalized hypergraph product [[882,24]] code;
\ref{code:hpc55-7938_578_16} is the~$10$-limited hypergraph product [[7938,578,16]] code;
\ref{code:haah8} is the~$8$-limited Haah's cubic [[1024,30]] code (see Appendix \ref{app:codes}).}
\label{fg:haah}
\end{figure}

In Fig.~\ref{fg:haah} we show the effect of the OSD-0 post-processing after the BP decoder for different QLDPC codes, including some known and new codes described in the next sections. As you can see, the gain of the BP-OSD  over the BP decoder in terms of WER for some codes is up to $5$ orders of magnitude (code~\ref{code:ghpc33-882_24}). However, for code~\ref{code:gbc-254_28} the difference between BP-OSD and BP is quite small. We think that this is because the column weight of the matrices $H_X$ and $H_Z$ is quite high (it is equal to~$5$), and hence the performance of the BP decoder is very good even without the post-processing. 

\begin{remark*}
It~is interesting to note that $8$-limited Haah's [[1024,30]] code, which has local stabilizers in 3D, also performs very well under the BP-OSD decoder. To the best of our knowledge, this is the first such demonstration on the depolarizing channel.   From our point of view, this observation implicitly suggests that some of Haah's cubic codes may have very good minimum distances. In~fact, even after very long runs of the BP-OSD decoder on this [[1024,30]] code, we have not found any non-degenerate codewords of weight less than $32$.
\end{remark*}

In Fig.~\ref{fg:OSD-order} we show the impact of the~OSD order $w$ on the WER performance of the~$8$-limited [[882,48,16]] code (Appendix \ref{app:codes}, code \ref{code:ghpc35-882_48}) under the BP-OSD decoder. We see that the WER performance in this case can be further improved by increasing the OSD order $w$. At the same time, from our experiments with the OSD post-processing on different QLDPC codes, we observed that in many cases the impact of the OSD order on the WER performance is quite small.

\begin{figure}
\begin{tikzpicture}
\pgfplotsset{major grid style={dashed}} 
\pgfplotsset{minor grid style={dotted}}
\pgfplotsset{
    tick label style={font=\footnotesize},
    label style={font=\footnotesize},
    legend style={font=\scriptsize},
}

\begin{loglogaxis}[
ylabel=WER,
xlabel=Physical error rate,
ymax=1,
width=\linewidth,
legend columns=2,
legend style={
	cells={anchor=west},
	legend pos= north west,
},
grid=both,
title=BP-OSD of different order $w$,
xtick={1e0, 1e-1, 1e-2},
extra x ticks={5e-2,2e-1},
	every axis y label/.style=
    {at={([xshift=-2pt]ticklabel cs:0.5)},rotate=90,anchor=center,font=\footnotesize},
]%
\addplot[red,mark=o] table[x=P,y=WER] {
P      WER
0.14   0.840336
0.13   0.518135
0.12   0.269542
0.11   0.0874891
0.1   0.0449236
0.09   0.0228102
0.08   0.0120034
0.07   0.0048895
0.06   0.00238209
0.05   0.00127979
0.04   0.000438333
0.03   0.000172334
};
\addlegendentry{no OSD}
\addplot[blue,mark=o] table[x=P,y=WER] {
P WER
0.14   0.8
0.13   0.490196
0.12   0.208768
0.11   0.0751315
0.1   0.0212359
0.09   0.012832
0.08   0.00459855
0.07   0.00173232
0.06   0.000633364
0.05   0.000180688
0.04   4.27447e-05
0.03   8.1182e-06
};
\addlegendentry{OSD-0}

\addplot[blue,mark=*] table[x=P,y=WER] {
P      WER
0.14   0.775194
0.13   0.46729
0.12   0.189753
0.11   0.0537057
0.1   0.0196078
0.09   0.00707864
0.08   0.00315597
0.07   0.00138702
0.06   0.000428399
0.05   9.5975e-05
0.04   2.74747e-05
0.03   4.72767e-06
};
\addlegendentry{OSD-4}

\addplot[black,mark=o] table[x=P,y=WER] {
P     WER
0.1   0.0111111
0.09   0.00423316
0.08   0.00149582
0.07   0.000568279
0.06   0.000192459
0.05   4.55689e-05
0.04   1.24895e-05
0.03   2.84818e-06
};
\addlegendentry{OSD-10}

\addplot[black,mark=*] table[x=P,y=WER] {
P      WER
0.14   0.740741
0.13   0.454545
0.12   0.160256
0.11   0.0401445
0.1   0.00881213
0.09   0.00262343
0.08   0.00111722
0.07   0.000342288
0.06   0.000124482
};
\addlegendentry{OSD-15}

\end{loglogaxis}
\end{tikzpicture}

\caption{The impact of the OSD order $w$ on the WER performance of the $8$-limited [[882,48,16]] code \ref{code:ghpc35-882_48} in Appendix~\ref{app:codes}.}
\label{fg:OSD-order}
\end{figure}

\subsection{Different post-processing algorithms}

In this section, the OSD post-processing algorithm is compared against the other known post-processing methods that also significantly improve the performance of the BP decoder. We assume that the probability distributions in the BP decoder are represented in terms of the log-likelihood ratios (LLRs). Below you can find a~brief description of these methods, where each method can be applied repeatedly after the BP decoder until all the parity-checks are satisfied or the maximal number of attempts $n_a$ is reached.   
\begin{itemize}
    \item {\bf Random perturbation} \cite{Poulin:2008}. If the syndrome is non-zero after the BP decoding, then we randomly choose an unsatisfied c-node and randomly change the~initial LLRs for all the v-nodes adjacent to this c-node. The main parameter of this method is the variance of the perturbation magnitude.
    After the changes are done, the BP decoder runs with the perturbed input LLRs. 
    \item {\bf Enhanced feedback} \cite{Wang:2012}. It is similar to the previous approach but the perturbations are not random and calculated using the previous BP output. If after the BP decoding the parity-checks are not satisfied, we randomly choose an~unsatisfied c-node. Then for all the~v-nodes adjacent to this c-node, we set the~initial LLRs using the BP decoder output for these nodes. After this is done the BP decoder runs with the perturbed input LLRs. 
    \item {\bf Matrix augmentation} \cite{Rigby:2019}. In this method instead of modification of the input LLRs, the parity-check matrix itself is modified by random duplication of some rows. The fraction~$\delta$ of the~duplicated rows is called the~\emph{augmentation
density}. Then a~new decoding attempt with the augmented matrix is performed. 
\end{itemize}

To compare the performance of all the described methods with the~BP-OSD decoder we use the $6$-limited $[[1270,28]]$ QLDPC code (\ref{code:ghpc33-1270_28} in Appendix \ref{app:codes}). This code belongs to the new class of codes described in Section~\ref{sc:ghp}. For all these methods we used $n_a=100$.
We can see in~Fig.~\ref{fig:ppcomp} that the OSD post-processing outperforms all the above-mentioned post-processing methods and also outperforms the~$4$-limited $[[1201,1,25]]$ surface code on the~MPS decoder from~\cite{Bravyi:2014}, which is almost optimal for this code. Let us note that all the other post-processing methods also have the WER gain that is more than $10^3$ over the BP decoder for code~\ref{code:ghpc33-1270_28}. In fact, we observed a~similar WER gain for many other codes from the class of $(3,6)$-regular CSS QLDPC codes with a~sufficiently large minimum distance. We think that this is mainly because the classical $(3,6)$-regular LDPC codes defined by the matrices $H_X$ and $H_Z$ of such CSS codes themselves have many harmful trapping sets. Thus these $(3,6)$-regular QLDPC codes additionally have a~lot of degenerate codewords of very low weight starting from $6$.

\begin{figure}
	\begin{tikzpicture}
	\pgfplotsset{major grid style={dashed}} 
	\pgfplotsset{minor grid style={dotted}}
	\pgfplotsset{
        tick label style={font=\footnotesize},
        label style={font=\footnotesize},
        legend style={font=\scriptsize},
        standard/.style={
            every axis y label/.style={at={(ticklabel cs:0.5)}, rotate=90,anchor=center,font=\footnotesize},
            every axis x label/.style={at={(ticklabel cs:0.5)}, anchor=center,font=\footnotesize}
        }
    }
	\begin{loglogaxis}[
	ylabel=WER,
	xlabel=Physical error rate,
	legend columns=2,
	legend style={
		cells={anchor=west},
		legend pos=north west,
	},
	every axis y label/.style=
    {at={(ticklabel cs:0.5)},rotate=90,anchor=center,font=\footnotesize},
	grid=both,
	title=BP post-processing algorithms,
	xtick={1e0, 1e-1, 1e-2},
	extra x ticks={5e-2,2e-1},
	xmax=0.2,
	xmin=0.018,
	enlarge y limits=.01,
	height=7.5cm,
	]%
	\addplot[black,mark=o] table[x=P,y=WER] {sims/res_ghpc33-1270_28_16.txt};
	\addlegendentry{Pure BP}

	\addplot[blue,mark=o] table[x=P,y=WER] {sims/res_ghpc33-1270_28_16-retry100-2.txt};
	\addlegendentry{Pert.}
	
	\addplot[blue,mark=*] table[x=P,y=WER] {sims/res_ghpc33-1270_28_16-retry100enh.txt};
	\addlegendentry{EFB}

	\addplot[green!50!black,mark=*] table[x=P,y=WER] {sims/res_ghpc33-1270_28_16-retry100aug.txt};
	\addlegendentry{Aug.}
	
	\addplot[black,mark=*] table[x=P,y=WER] {sims/res_ghpc33-1270_28_16-osd.txt};
	\addlegendentry{OSD-0}

	\addplot[red,mark=square*] table[x=P,y=WER] {sims/toric_d25deep.txt};
	\addlegendentry{Surf. MWM}

	\addplot[red,mark=*] table[x=P,y=WER] {sims/surf25-mps.txt};
	\addlegendentry{Surf. MPS}
		
	\end{loglogaxis}
	\end{tikzpicture}
	
	\caption{The WER performance of different post-processing algorithms for the BP decoder on the~\mbox{$6$-limited} [[1270,28]] QLDPC code (\ref{code:ghpc33-1270_28} in Appendix~\ref{app:codes}). The red curves are for the $4$-limited [[1201,1,25]] surface code under the~minimum weight matching (MWM) and the~MPS-based decoders~\cite{Edmonds:1965, Bravyi:2014}.}\label{fig:ppcomp}
\end{figure}

\section{\label{sc:gbc}New generalized bicycle codes}
\subsection{Ansatz with commuting matrices}
The commutativity conditions such as (\ref{eq:comm-cond}) and (\ref{eq:comm-cond-CSS}) are a~serious obstacle to designing good QLDPC codes using random-like constructions similar to the constructions used for classical LDPC codes. Thus it makes sense to consider large families of matrices of some particular form called \emph{ansatz}, where the commutativity conditions are always satisfied. One such quite general ansatz for CSS codes was proposed in~\cite{Kovalev&Pryadko:HBP:2013} as a generalization of the bicycle QLDPC codes~\cite{Mackay:2004}. Let us briefly remind this ansatz. Consider two commuting binary $n\times n$ matrices $A$ and $B$, i.e., $AB = BA$. Let 
\begin{equation}\label{eq:comm-ansatz}
H_X = [A, B] \text{ and } H_Z = [B^\T, A^\T]. 
\end{equation}
Then we see that $H_X H_Z^\T = AB + BA = \zm$, and the commutativity condition (\ref{eq:comm-cond-CSS}) is always satisfied.  
It was proposed in~\cite{Kovalev&Pryadko:HBP:2013} to use binary circulant matrices $A$ and $B$ since they always commute. The corresponding class of codes is called the \emph{generalized bicycle (GB) codes}, where the~bicycle codes~\cite{Mackay:2004} are obtained as a special case when $B=A^\T$. 

\subsection{Ring of circulants}\label{sc:circ}
Let us recall that an $\ell\times\ell$ circulant matrix $\mat{A}$ over $\F_q$ takes the form
\[
\mat{A} =
\begin{pmatrix}
a_{0} & a_{l-1} & \ldots & a_{1}\\
a_{1} & a_{0} & \ldots & a_{2}\\
\hdotsfor{4}\\
a_{\ell-1} & a_{\ell-2} & \ldots & a_{0}\\
\end{pmatrix},
\]
where $a_0,\ldots,a_{\ell-1}\in \F_q$. It is readily seen that the matrix $\mat{A}$ can be represented in the form
\[\mat{A} = a_0\mat{I} + a_1\mat{P} + \dots a_{\ell-1}\mat{P}^{\ell-1},\]
where $\mat{I}$ is the $\ell\times\ell$ identity matrix and
\[
\mat{P} =
\scalebox{0.9}{$
\begin{pmatrix}
0 & 0 & \ldots & 1\\
1 & 0 & \ldots & 0\\
0 & 1 & \ldots & 0\\
\hdotsfor{4}\\
0 & 0 & \ldots & 0\\
\end{pmatrix}$
}
\]
is the $\ell\times\ell$ permutation matrix representing the \emph{right} cyclic shift by \emph{one} position. Since $\mat{P}^\ell = \mat{I}$, we see that the ring of all $\ell\times\ell$ circulant matrices over $\F_q$ is isomorphic to the ring $\RC{q}{\ell}=\F_q[x]/(x^\ell-1)$ of polynomials over $\F_q$ modulo the polynomial $x^\ell - 1$. 

Hence the circulant matrix $\mat{A}$ can be uniquely represented by the polynomial
$a(x)=a_0 + a_1 x + \dots + a_{\ell-1} x^{\ell-1}$
and the product $\mat{C}=\mat{A}\mat{B}$ of two circulant matrices represented by polynomials $a(x),b(x)\in R_\ell$ corresponds to the polynomial
\begin{equation}\label{eq:cycl-conv}
c(x) = a(x)b(x) \mod x^\ell -1
\end{equation}
which is called the \emph{cyclic convolution} of $a(x)$ and $b(x)$. Likewise, if we want to find a matrix-vector product ${\vec{c}=\mat{A}\vec{b}}$, where $\vec{b}=(b_0,\dots,b_{\ell-1})$ and $\vec{c}=(c_0,\dots,c_{\ell-1})$ are (column) vectors corresponding to $b(x)$ and $c(x)$, we can also use the cyclic convolution (\ref{eq:cycl-conv}).

\subsection{Dimension of generalized bicycle codes}
As we saw before, to define two binary circulant $\ell\times\ell$ matrices $A$ and $B$ we need to provide two binary polynomials $a(x), b(x)\in \RC{2}{\ell}$. The dimension\footnote{Let us point out that this dimension formula was given in the paper~\cite[Theorem~2]{Kovalev&Pryadko:HBP:2013} in a slightly more complex form. A similar formula was proved only in the special case of single
generator codes.} $k$ of the corresponding CSS $[[2\ell, k]]$ code is given by the following proposition.
\begin{proposition}\label{pr:gbc-dim}
The dimension $k$ of the the generalized bicycle $[[2\ell, k]]$ code defined by $a(x), b(x)\in\RC{2}{\ell}$ is given by the formula:
\begin{equation}\label{eq:gbc-dim}
    k=2\deg g(x),
\end{equation}
where $g(x) = \gcd(a(x), b(x), x^\ell - 1)$. 
\end{proposition}

To prove this formula we show that $\rk H_X = \rk H_Z = n- \deg g(x)$ and use CSS dimension formula (\ref{eq:CSS-dim}).   

Let us first find the rank of the matrix $H_X = [A, B]$, which is equal to the dimension of its column space. It is easy to see that the column space of $H_X$ (called its \emph{syndrome space}) is equal to the following set:
\[
\{A u + B v \mid u, v\in \F_2^\ell\}.
\]
Using the described above polynomial representation of column vectors and circulant matrices we can consider this set as the following set of polynomials from $\RC{2}{\ell}$:
\[
\{a(x) u(x) + b(x) v(x) \mid u(x), v(x)\in \RC{2}{\ell}\}.
\]
It is easy to verify that this set is the~principal\footnote{An ideal $I$ in a ring $R$ is \emph{principal} if $I = \{au \mid u\in R\}$ for some $a \in R$.}  ideal of the ring $\RC{2}{\ell}$ generated by $g(x)$. Hence it is the cyclic code $\cC_g$ with generator polynomial $g(x)$, and we proved that $\rk H_X = \dim \cC_g = n -\deg g(x)$. We call this cyclic code $\cC_g$ the \emph{syndrome code} of $H_X$ since its codewords are precisely the polynomial representations of the syndrome space of $H_X$. 

To complete the proof we also need to show that $\rk H_Z = n - \deg g(x)$. Using similar arguments as above we can consider the syndrome code for $H_Z$ and show that it is generated by the ``transposed'' polynomial $g^*(x) = g(x^{-1})$. Though the codes generated by the polynomials $g(x)$ and $g^*(x)$ may  differ, they always have the same dimension since the corresponding circulant matrices $G$ and $G^\T$ have the same rank. Hence we also proved that $\rk H_Z = n - \deg g(x)$, and the proof of formula~(\ref{eq:gbc-dim}) is complete.

\subsection{Construction methods}

The proof of Proposition~\ref{pr:gbc-dim} gives us also some valuable information on how to find generalized bicycle codes of high dimension. If we fix the circulant size $\ell$ then all possible dimensions $k$ of the generalized bicycle codes with this circulant size are characterized by the degrees of all possible factors of the polynomial $x^\ell - 1$. Indeed, for each factor $g(x)$ of the polynomial $x^\ell - 1$ we can always choose polynomials $a(x), b(x)\in\RC{2}{\ell}$ such that: 
\begin{equation}\label{eq:a-b-cond}
    a(x)\mod g(x) = b(x) \mod g(x) = 0
\end{equation}
since these polynomials are just two codewords from the cyclic code~$\cC_g$ generated by $g(x)$, which we called the syndrome code of $H_X$. To produce a~$w$-limited QLDPC we just need to find low weight polynomials $a(x)$ and $b(x)$ that are the codewords of $\cC_g$. In practice, this can be accomplished by several methods. If the circulant size $\ell$ is relatively small, we can find $a(x), b(x)$ by an exhaustive search over all polynomials of the given weight. 

Another alternative is to generate random polynomials of a~given weight from $\RC{2}{\ell}$ until we find a~pair of polynomials that satisfies condition~(\ref{eq:a-b-cond}). Since the probability that a~random polynomial of a~given weight belongs to $\cC_g$ is equal approximately to $2^{-\deg g(x)}$; then if we test more then $2^{\deg g(x)}$ random polynomials we will find a polynomial from $\cC_g$ with high probability. 

When we find a pair of polynomials $a(x), b(x)$ we also need to check that the corresponding code has good error correcting performance. This can be done by a simulation of the corresponding generalized bicycle code.

All the generalized bicycle codes from Appendix~\ref{app:codes} were found by the described above methods.   

\subsection{GB codes with syndrome protection}

Another important observation, made in the proof of Proposition~\ref{pr:gbc-dim}, is that the syndrome codes of the parity-check matrices $H_X$ and $H_Z$ are the cyclic codes with the generator polynomials $g(x)$ and $g^*(x)$, respectively. Let us mention that the syndrome code of a parity-check matrix is precisely the set of all possible syndromes for it. Hence if we use generator polynomials $g(x)$ and $g^*(x)$ that define cyclic codes with minimum distance $d$, we see that the syndromes for matrices $H_X$ and $H_Z$ are protected by these cyclic codes. Since the syndrome measurements for quantum codes are performed by faulty hardware, some additional protection of the syndromes may be used to improve the reliability of the syndrome measurements~\cite{Bombin:syn:2013, Fujiwara:syn:2014}.  Let us also mention that the polynomials $g(x)$ and $g^*(x)$ always produce cyclic codes $\cC_g$ and $\cC_{g^*}$ with the same  minimum distance since the ``transpose'' map 
\[
    c(x)=\sum_{i=0}^{\ell-1} c_i x^i\mapsto c^*(x) = \sum_{i=0}^{\ell-1} c_i x^{\ell - i} \mod x^\ell - 1
\]
is an automorphism of the ring $\RC{2}{\ell}$ that respects the weight of the polynomials, and therefore we have that $\cC_{g^*} = \{c^*(x)  \mid c(x)\in \cC_g \}$.

Hence we can use any cyclic code with generator polynomial $g(x)$ and minimum distance $d$ to construct a~CSS code, where the syndromes for $H_X$ and $H_Z$ are protected by cyclic codes of minimum distance $d$. It~is important to note that since $a(x), b(x) \in \cC_g$, the weight of the polynomials $a(x), b(x)$ can not be smaller than this minimum distance $d$.

\subsection{Comparison with other codes}\label{sc:GB-constr}

\begin{figure}
	\begin{tikzpicture}
	\pgfplotsset{major grid style={dashed}} 
	\pgfplotsset{minor grid style={dotted}}
	\pgfplotsset{
        tick label style={font=\footnotesize},
        label style={font=\footnotesize},
        legend style={font=\scriptsize},
    }
	\begin{loglogaxis}[
	ylabel=WER,
	xlabel=Physical error rate,
	legend style={
		cells={anchor=west},
		legend pos=north west,
	},
	grid=both,
	ymax=1,
	title=HMP vs GB (BP-OSD-10),
	every axis y label/.style=
    {at={([xshift=-2pt]ticklabel cs:0.5)},rotate=90,anchor=center,font=\footnotesize},
	]

	\addplot table[x=P,y=WER] {sims/res_hmp-49_1_9-osd10.txt};
    \addlegendentry{HMP [[49,1,9]]}

    \addplot table[x=P,y=WER] {sims/res_gbc44-48_6_8-osd10.txt};
    \addlegendentry{GB [[48,6,8]]}
    
    \addplot table[x=P,y=WER] {sims/res_gbc44-46_2_9-osd10.txt};
    \addlegendentry{GB [[46,2,9]]}
    
	\end{loglogaxis}
	\end{tikzpicture}
	
	\caption{The WER performance of the $8$-limited generalized bicycle (GB) codes (\ref{code:gbc-48_6_8} and \ref{code:gbc-46_2_9} in Appendix \ref{app:codes}) and the $8$-limited $[[49,1,9]]$ homological product (HMP) code from \cite{Bravyi:HMP:2014} under the BP decoder with the OSD-10 post-processing.}
	\label{fg:hmp}
\end{figure}

In this subsection, we consider several new examples of generalized bicycle codes and compare their performance under the BP-OSD decoder against some other already known QLDPC codes.

\begin{example}
Let us consider the primitive narrow-sense BCH $[127, 14, 5]$ code $\cC_g$ with the generator polynomial: \[
g(x) = (x^7 + x + 1)\cdot (x^7 + x^5 + x^3 + x + 1).
\]
If we set $a(x) = 1+x^{15}+x^{20}+x^{28}+x^{66}$ and $b(x) = 1+x^{58}+x^{59}+x^{100}+x^{121}$, then we obtain the $10$-limited generalized bicycle $[[254, 28]]$ code. Its minimum distance is not available, but the performance of this code (see Fig.~\ref{fg:haah}, code~\ref{code:gbc-254_28}) is almost the same as the performance of the $10$-limited hypergraph product $[[7938,578,16]]$ code\footnote{The definition of these codes is given in Section~\ref{sc:ghp}.} obtained from the~two identical circulant parity-check matrices $H$ of the cyclic  code $[63, 17, 16]$ code defined by the polynomial $h(x) = 1 + x^{3} + x^{34} + x^{41} + x^{57}$. This particular code was chosen in order to match the high rate of the $[[254, 28]]$ code. It is important to note that both codes do not have $4$-cycles in matrices $H_X$ and $H_Z$, and they have the same weight of stabilizers. It is also interesting that the performance of the $[[254, 28]]$ code is almost the same even under the classical BP decoder without OSD post-processing. The reason of such good performance with the BP decoder is not fully understood. One of the possible explanations is related with the trapping set structure of the  $[[254, 28]]$ code. Since its syndrome code has minimum distance $5$, it can not have $(a, b)$ trapping sets\footnote{An \emph{$(a, b)$ trapping set} or a~\emph{near-codeword} for a parity-check  matrix~$H$ is a vector $v$ of weight $a$ such that the corresponding syndrome $s = H v$ has weight $b$.} with $b < 5$. It is very well known~\cite{Richardson:error-floor:2003} that $(a, b)$ trapping sets with small $a$ and $b$ may greatly decrease the performance of the BP decoder. And the $[[254, 28]]$ can not have the most harmful trapping sets.
\end{example}

\begin{figure}
\begin{tikzpicture}
\pgfplotsset{major grid style={dashed}} 
\pgfplotsset{minor grid style={dotted}}
	\pgfplotsset{
        tick label style={font=\footnotesize},
        label style={font=\footnotesize},
        legend style={font=\scriptsize},
    }
\begin{loglogaxis}[
title=Bicycle vs Generalized Bicycle (NBP and BP),
ylabel=Logical error rate,
xlabel=Physical error rate,
ymax=1,
ymin=5e-7,
legend style={
	cells={anchor=west},
	legend pos=north west,
},
grid=both,
	every axis y label/.style=
    {at={([xshift=-3pt]ticklabel cs:0.5)},rotate=90,anchor=center,font=\footnotesize},
]%
\addplot table[x=P,y=BER] {sims/res_bc-256_32-BP.txt};
\addlegendentry{[[256,32]] BP (binary)}
\addplot table[x=P,y=BER] {sims/res_bc-256_32-NBP.txt};
\addlegendentry{[[256,32]] NBP}
\addplot table[x=P,y=WER] {sims/res_gbc55-126_28_8-bin.txt};
\addlegendentry{[[126,28]] BP (binary)}
\addplot table[x=P,y=WER] {sims/res_gbc55-126_28_8.txt};
\addlegendentry{[[126,28]] BP}

\end{loglogaxis}
\end{tikzpicture}
\caption{The logical error rate performance of the $16$-limited $[[256, 32]]$ bicycle code under the neural BP (NBP) decoder  from~\cite{Liu:2019} and the $10$-limited $[[126,28,8]]$ generalized bicycle code (\ref{code:gbc-126_28} in Appendix~\ref{app:codes}) under the BP decoder (binary and non-binary) \emph{without} the OSD post-processing.}
\label{fg:neural}
\end{figure}

\begin{example}
Let us consider the cyclic $[63, 14, 5]$ code with the generator polynomial
\[
    g(x) = (x^2 + x + 1) \cdot (x^6 + x^5 + 1) \cdot (x^6 + x^5 + x^4 + x + 1)
\]
If we set $a(x) = 1 + x + x^{14} + x^{16} + x^{22}$, $b(x) = 1 + x^{3} + x^{13} + x^{20} + x^{42}$ we obtain the $10$-limited generalized bicycle $[[126, 28]]$ code.
Its performance with the standard BP decoder (binary and non-binary) is shown in Fig.~\ref{fg:neural}, code~\ref{code:gbc-126_28}.
We compared its performance with the performance of the neural BP decoder for the~bicycle $[[256, 32]]$ code constructed in~\cite{Liu:2019}. 
Such a~big difference in the performance is mostly because the QLDPC $[[256, 32]]$ code used in~\cite{Liu:2019} has a~small minimum distance compared to the $[[126,28]]$ code, which minimum distance is $8$. This minimum distance was found by an~exhaustive search (see Appendix~\ref{app:codes}). Another possible reason is that the neural BP decoder proposed in~\cite{Liu:2019} was based on the binary BP, which usually has worse performance than its non-binary version.
\end{example}

\begin{figure}
	\begin{tikzpicture}
	\pgfplotsset{major grid style={dashed}} 
	\pgfplotsset{minor grid style={dotted}}
	\pgfplotsset{
        tick label style={font=\footnotesize},
        label style={font=\footnotesize},
        legend style={font=\scriptsize},
    }
	\begin{loglogaxis}[
	ylabel=WER,
	xlabel=Physical error rate,
	legend style={
		cells={anchor=west},
		legend pos=north west,
	},
	grid=both,
	title=HB vs GB (BP-OSD-10),
	xtick={1e0, 1e-1, 1e-2},
	extra x ticks={5e-2,2e-1},
	enlarge y limits=.01,
	ymax=1,
	every axis y label/.style=
    {at={([xshift=-2pt]ticklabel cs:0.5)},rotate=90,anchor=center,font=\footnotesize},
	]%

	\addplot table[x=P,y=WER] {sims/res_hbc-900_50_14.txt};
    \addlegendentry{HB [[900,50,14]]}

    \addplot table[x=P,y=WER] {
    P      WER
    0.14   0.787402
    0.13   0.645161
    0.12   0.37594
    0.11   0.294118
    0.1   0.157978
    0.09   0.0665336
    0.08   0.0254842
    0.07   0.0074217
    0.06   0.00172697
    0.05   0.000261949
    0.04   3.48687e-05
    0.03   3.39816e-06
    0.02   1.39792e-07
    };
    \addlegendentry{GB [[180,10, $d\ge 15$]]}
    \addplot table[x=P,y=WER] {sims/res_gbc44-900_50_15-g4-osd.txt};
    \addlegendentry{GB [[900,50,15]]}
	\end{loglogaxis}
	\end{tikzpicture}
	
	\caption{The WER performance of the $8$-limited hyperbicycle (HB) [[900,50,14]] code and two $8$-limited generalized bicycle (GB) codes  (\ref{code:gbc44-180_10} and \ref{code:gbc44-900_50} in Appendix~\ref{app:codes}) under the BP-OSD-10 decoder.}
	\label{fg:hbc}
\end{figure}

\begin{example}
In this example we constructed two very small $8$-limited generalized bicycle codes (the $[[48, 6, 8]]$ code~\ref{code:gbc-48_6_8} and the $[[46,2,9]]$ code~\ref{code:gbc-46_2_9}, see Appendix~\ref{app:codes}). We compared their performance (see Fig.~\ref{fg:hmp}) with the performance of an~$8$-limited $[[49, 1, 9]]$ homological product (HMP) code from~\cite{Bravyi:HMP:2014} using the BP with OSD-like post-processing. As we can see the performance of the newly constructed codes is similar to the $[[49, 1, 9]]$ code. At the same time, their rates are higher. 
\end{example}

\begin{example}
In Fig.~\ref{fg:hbc} we compared the performance of the $8$-limited hyperbicycle [[900, 50, 14]] code from~\cite{Kovalev&Pryadko:2012, Kovalev&Pryadko:HBP:2013} with two new $8$-limited generalized bicycle codes (\ref{code:gbc44-180_10} and \ref{code:gbc44-900_50} in Appendix~\ref{app:codes}). We can see that the performance of the generalized bicycle $[[180,10,d]]$ code, $15\le d \le 18$, is similar to the hyperbicycle [[900, 50, 14]] code. At the same time, it has the same weight of stabilizers, and its code length is $5$ times smaller.
\end{example}

\section{Generalization of HP codes}\label{sc:ghp}

\subsection{Hypergraph product (HP) codes}

In this section, we propose a~new generalization of hypergraph product codes~\cite{Tillich&Zemor:2009} in the case when one of the parity-check matrices in the product is square. Let us first remind the definition of these codes in a matrix form~\cite{Kovalev&Pryadko:2012}. Suppose we have an $[n_a, k_a, d_a]$ linear code $\cC_a$ and an $[n_b, k_b, d_b]$ linear code $\cC_b$ defined by parity-check matrices\footnote{The parity-check matrices are not necessary full rank.} $a\in\Mat_{m_a\times n_a}(\F_2)$ and $b\in\Mat_{m_b\times n_b}(\F_2)$ respectively. Then the \emph{hypergraph product} code is the CSS $[[N, K, d]]$ code with $H_X = (a\otimes I_{m_b}, I_{m_a}\otimes b)$ and $H_Z = (I_{n_a}\otimes b^\T, a^\T\otimes I_{n_b})$, where $N = n_a m_b + n_b m_a$, $K = 2k_a k_b - k_a(n_b - m_b) - k_b(n_a - m_a)$. As it was shown in~\cite{Tillich&Zemor:2009}, the minimum distance $d$ of the hypergraph product code $\cC$ satisfies the following lower bound:
\[
d \ge \min(d_a, d_b, d_a^\T, d_b^\T),
\]
where $d_a^\T$ and $d_b^\T$ are the minimal distances of the ``transposed'' codes $C_a^\T$ and $C_b^\T$ defined by the parity-check matrices $a^\T$ and $b^\T$ respectively. It is important to note that if the matrices $a$ and $b$ are $w$-limited, then the corresponding CSS code $\cC$ is $2w$-limited. Hence, using known asymptotically good families of classical LDPC codes with $(w_c,w_r)$-limited parity check-matrices, it is possible~\cite{Tillich&Zemor:2009} to construct $w$-limited CSS codes with asymptotically non-zero rate and $d=\Theta(\sqrt{n})$ as $n\to\infty$. In~\cite{Kovalev&Pryadko:2012} the hypergraph product construction was further improved, and it was shown that one can construct  good hypergraph product codes using square parity-check matrices $a$ and $b$. In fact, many of the best-known small-length hypergraph product codes are constructed using square parity check matrices of cyclic codes (see~\cite{Kovalev&Pryadko:2012, Kovalev&Pryadko:HBP:2013}). In~\cite{Kovalev&Pryadko:HBP:2013}  hyperbicycle CSS codes, which generalize both generalized bicycle and hypergraph product codes, were proposed. Here we consider another generalization of hypergraph product codes where the matrix $b$ is square.

\subsection{Generalized hypergraph product codes}

In what follows by a~\emph{ring} we always mean a~ring with identity.
Let $R$ be a ring. We denote the ring of all $m\times n$ matrices over $R$ by $\Mat_{m\times n}(R)$ or by $\Mat_{n}(R)$ when $m=n$. If $R$ is the~ring of $\ell \times \ell$ matrices over some field $\F$ we identify the elements of $\Mat_{m\times n}(R)$ with the corresponding block matrices from $\Mat_{m\ell\times n\ell}(\F)$.

Consider a binary matrix $\mat{b}\in\Mat_\ell(\F_2)$. We say that the matrix~$b$ and a~ring $R\subseteq \Mat_\ell(\F_2)$ \emph{commute} if all matrices from $R$ commute with $b$. 

\begin{example}\label{ex:hgp-ring}
Consider $b\in\Mat_\ell(\F_2)$ and $R=\{\zm,I\}$, where $\zm$ and $I$ are the zero and the identity matrices from $\Mat_\ell(\F_2)$ respectively; then $b$ and $R$ always commute.
\end{example}

\begin{example}\label{ex:ghgp-ring}
Let $b$ be a binary circulant matrix  and $R$ be the ring of all binary circulant matrices of the same size; then $b$ and $R$ always commute. 
\end{example}

Suppose that a matrix $b\in\Mat_\ell(\F_2)$ and a ring $R\subseteq \Mat_\ell(\F_2)$ commute. Consider a matrix $A   = (a_{i j})_{m\times n} \in\Mat_{m\times n}(R)$. We denote by $\cC(A, b)$ the CSS code called a~\emph{generalized hypergraph product} (\emph{GHP}) code with the following parity-check matrices\footnote{Let us warn the reader that we understand $H_X$ and $H_Z$ as the corresponding binary block matrices (not as matrices over $\Mat_\ell(\F_2)$).}:
\begin{equation}\label{eq:gen-ansatz}
    H_X = [A, bI_m], H_Z = [b^\T I_n, A^*],
\end{equation}
where $A^*=(a_{j i}^T)_{n\times m}$, and  $I_m, I_n$ are the identity matrices over $R$ of size $m$ and $n$ respectively.
The correctness of this definition follows from the following:
\[
    H_X H_Z^\T = [A, bI_m]
    \begin{bmatrix}
    I_n b\\
    A \\
    \end{bmatrix}
    = A b + b A = \zm.
\]

The code length $N$ of the CSS code $\cC(A, b)$ is given by $N = (m + n)\ell$. We will show later how to find the dimension $K$ of the code $\cC(A, b)$ in a special case.

One can easily verify that if we take a matrix $b\in\Mat_\ell(\F_2)$ and the~ring $R=\{\zm,I\}$ (as in Example~\ref{ex:hgp-ring}) then the CSS code $\cC(A, b)$ is the hypergraph product code defined by the binary $m\times n$ matrix $\tilde{a}=(\tilde{a}_{i j})_{m\times n}$ and the binary $\ell\times \ell$ matrix $b$, where for all $i\in [m]$, $j \in [n]$ we have:
\[
    \tilde{a}_{i j} = 
    \begin{cases}
        0, & \text{if $a_{i j} = \zm$;} \\
        1, & \text{if $a_{i j} = I$.}
    \end{cases}
\]

We can also see that the ansatz with two commuting matrices $A$ and $B$ given in~(\ref{eq:comm-ansatz}) is also a special case of the new ansatz described in~(\ref{eq:gen-ansatz}), where the matrix $A$ is a~$1\times 1$ block matrix.

\subsection{Quasi-cyclic generalized hypergraph product codes}

In this subsection, we describe quasi-cyclic (QC) GHP codes. In fact, it can be shown that this particular subclass of GHP codes is equivalent (up to some permutation of qubits) to a~special case of hyperbicycle codes from~\cite[Eq.~(19), $\chi=r_2=n_2=1$]{Kovalev&Pryadko:HBP:2013}. However, we believe that the concise language of polynomial matrices adopted in the current work is much more suitable for our examples, which are essentially sparse random QC~GHP codes subject to some additional constrains.

Now, let us take $b$ and $R$ as in Example~\ref{ex:ghgp-ring}. Hence $b$ is a~binary circulant matrix, and $R$ is the ring of all binary circulant matrices of the same size. In this case, the matrices $H_X$ and $H_Z$ defined by~(\ref{eq:gen-ansatz}) are block matrices, where each block is a~binary circulant matrix of size~$\ell$. Such matrices are called \emph{quasi-cyclic}. Let us note that quasi-cyclic (QC) matrices are well known in classical coding theory. In fact, most of the best-known practical classical LDPC codes have QC parity-check matrices.  
We will show in this section how to find the dimension of generalized hypergraph product codes defined by~(\ref{eq:gen-ansatz}). For simplicity, we consider here only the case when the circulant size $\ell$ is odd. Before we can provide the formula for the dimension we need some supplementary definitions from algebra.

Here we adopt the polynomial representation of QC matrices used in~\cite{Lally:2001, Smarandache:2012}. 
For any polynomial $p(x)\in\F_q[x]$ of degree $d$ we consider the ring $\F_q[x]/(p(x))$ of  polynomials ${f_0 + f_1 x + \dots + f_{d-1}x^{d-1}\in \F_q[x]}$ with addition and multiplication modulo $p(x)$. By $\F_q^d$ we denote the \mbox{$d$-dimensional} space of the $d\times 1$ column vectors over~$\F_q$.  We also identify an~element $f(x)\in\F_q[x]/(p(x))$ with the corresponding column vector $\vec{f}=(f_0, \ldots, f_{d-1})\in \F_q^d$. 

Let us recall that by $\RC{2}{\ell}$ we denote the ring of circulants $\F_2[x]/(x^\ell - 1)$. We use the standard identification of the circulant $\ell\times\ell$ matrices over $\F_2$ with the elements of the ring $\RC{2}{\ell}$ (see Subsection~\ref{sc:circ}), where a~column vector $\vec{a}\in \RC{2}{\ell}$ corresponds to the circulant matrix with the first column equal to $\vec{a}$. Using this identification we can consider an $m\ell\times n\ell$ QC matrix over $\F_2$ of circulant size $\ell$ as an $m\times n$ matrix over the ring $\RC{2}{\ell}$. We also consider $n\times 1$ column vectors over $\RC{2}{\ell}$ as $n\ell\times 1$ column vectors over $\F_2$. Given the above identification we consider multiplication of an $m\ell\times n\ell$ QC matrix by an $n\ell\times 1$ column vector over $\F_2$ as multiplication of an $m\times n$ matrix by an $n\times 1$ column vector over $\RC{2}{\ell}$.

The algebraic structure of the ring $\RC{2}{\ell}$ is very well studied in the literature (see Appendix~\ref{app:alg} for further details). Since we consider the case when $\ell$ is odd, the  polynomial $x^\ell - 1$ factors into a product of  irreducible polynomials over $\F_2$:
\begin{equation}\label{eq:factors-co-prime1}
x^\ell - 1  = f_1(x)\cdots f_s(x).
\end{equation}
Hence the ring $\RC{2}{\ell}$ is isomorphic to the direct product of finite fields:  
\begin{equation}\label{eq:isom-co-prime1}
\RC{2}{\ell} \cong F_1\times \dots \times F_s,
\end{equation}
where the field $F_i = \F_2[x]/\bigl(f_i(x)\bigr)$ has the size $2^{d_i}$;  $d_i = \deg f_i(x)$,  $i\in [s]$. 
Let us consider the maps ${\varphi_i\colon \RC{2}{\ell}\to F_i}$ given by the formula: $$\varphi_i\colon u(x) \mapsto u(x) \mod f_i(x), i\in [s].$$
We also naturally extend this map to any vectors and matrices over $\RC{2}{\ell}$.
The following lemma is the key to the dimension formula for the QC generalized hypergraph product codes.

\begin{lemma}\label{lm:key-dim}
Let $A\in \Mat_{m\times n}(\RC{2}{\ell})$ be a binary QC matrix. Then its rank \textup{(}over $\F_2$\textup{)} is given by:
\[
    \rk_{\F_2} A = \sum_{i=1}^{s} d_i \rk_{F_i}  \varphi_i(A)
\]
\end{lemma}
\begin{proof}
The lemma easily follows from the isomorphism shown in~(\ref{eq:isom-co-prime1}) between $\RC{2}{\ell}$ and the direct product of the fields $F_1,\dots,F_s$. Indeed, let $r_i = \rk_{F_i}\varphi_{i}(A)$, $i\in[s]$.  If~$A_1,\dots,A_n$ are the columns of $A$, then each vector $v$ that belongs to the column space of the matrix $A$ can be represented as follows:
\[
    v = u_1 A_1 + \dots + u_n A_n,
\]
where $u_1,\dots,u_n\in\RC{2}{\ell}$.
Therefore we see that the cardinality of the column space of the non-binary matrix $\varphi_i(A)$ over the field $F_i$ is equal to $(2^{d_i})^{r_i} = 2^{d_i r_i}$. Hence using isomorphism~(\ref{eq:isom-co-prime1}) we conclude that the number of different vectors in the column space of the matrix $A$ is equal to 
$$\prod_{i=1}^{s} 2^{d_i r_i} = 2^{\sum_{i=1}^{s} d_i r_i},$$
and we proved that $\rk_{\F_2} A = \sum_{i=1}^{s} d_i r_i$.
\end{proof}

The following proposition provides the formula for the dimension of the  QC CSS code $\cC(A, b)$ if $\ell$ is odd.
\begin{proposition}\label{th:gbc-dim}
Let $\mat{b(x)}\in \RC{2}{\ell}$,  $A\in\Mat_{m\times n}(\RC{2}{\ell})$, where $\ell$ is odd. Let $g(x) = \gcd(b(x), x^\ell - 1) = \prod_{i\in S} f_i(x)$, $S \subseteq [s]$, where $f_i(x)$ are some irreducible polynomials from~\textup{(}\ref{eq:factors-co-prime1}\textup{)}, and $F_i = \F_2[x]/(f_i(x))$, $i\in S$, are the corresponding finite fields. Then $\cC(A, b)$ is a~CSS $[[N, K]]$ code, where $N=(m+n)\ell$ and 
$$K = \sum_{i \in S} \deg f_i(x) (m + n - 2\rk_{F_i} \varphi_i(A)).$$
\end{proposition}
\begin{proof}
The proof idea is the following.
Let us mention that $\varphi_i(H_X) = [\varphi_i(A), \zm]$ for all $i\in S$. Hence we have $\rk_{F_i} \varphi_i(H_X) = \rk_{F_i} \varphi_i(A)$, for all $i\in S$. At the same time for ${i \in [s]\setminus S}$, we obtain 
$$\rk_{F_i} \varphi_i(H_X) = \rk_{F_i} [\varphi_i(A), \varphi_i(b) I_m] = m,$$ 
since for all ${i \in [s]\setminus S}$ we have that $\varphi_i(b) \ne 0$ and therefore the non-binary matrix $\varphi_i(H_X)$ is full rank. Hence by applying Lemma~\ref{lm:key-dim} to the matrix $H_X$ we have:
\begin{align*}
    m\ell - \rk H_X &= \sum_{i=1}^{s} \deg f_i(x) (m - \rk_{F_i} \varphi_i(H_X))\\
    &= \sum_{i\in S} \deg f_i(x) (m - \rk_{F_i} \varphi_i(A)).  
\end{align*}

To complete the proof we need also to find the rank of the matrix $H_Z$. It is easier to find the rank of the matrix $H'_Z$ obtained from $H_Z$ by the application of the ``transpose'' map $u \mapsto u^*$ to each element. Since the transpose map is an automorphism on the ring $\RC{2}{\ell}$ we see that the number of vectors in the row space of $H_Z$ and the row space of $ H'_Z$ is  the same. Hence $\rk H_Z = \rk  H'_Z$. The rank of the matrix~$H'_Z = [b I_n, A^T]$, can be found in the same way as for the matrix $H_X$:

\begin{align*}
    n\ell - \rk  H'_Z &= \sum_{i=1}^{s} \deg f_i(x) (n - \rk_{F_i} \varphi_i( H'_Z))\\
    &= \sum_{i\in S} \deg f_i(x) (n - \rk_{F_i} \varphi_i(A^\T))\\  
    &= \sum_{i\in S} \deg f_i(x) (n - \rk_{F_i} \varphi_i(A)).  
\end{align*}

Now we apply formula~(\ref{eq:CSS-dim}) for the CSS dimension and obtain: 
\begin{align*}
K &= N - \rk H_X - \rk H_Z \\
&= (m\ell - \rk H_X) + (n\ell - \rk  H'_Z) \\
&= \sum_{i \in S} \deg f_i(x) (m + n - 2\rk_{F_i} \varphi_i(A)).
\end{align*}
This concludes the proof.
\end{proof}

\begin{figure}
	\begin{tikzpicture}
	\pgfplotsset{major grid style={dashed}} 
	\pgfplotsset{minor grid style={dotted}}
	\pgfplotsset{
        tick label style={font=\footnotesize},
        label style={font=\footnotesize},
        legend style={font=\scriptsize},
    }
	\begin{loglogaxis}[
	ylabel=WER,
	xlabel=Physical error rate,
	legend style={
		cells={anchor=west},
		legend pos=north west,
	},
	grid=both,
	title=GHP vs HP (BP and BP-OSD-10),
    every axis x label/.style=
    {at={(ticklabel cs:0.5)},anchor=center,font=\footnotesize},
    enlarge y limits=.01,
    	every axis y label/.style=
    {at={([xshift=-3pt]ticklabel cs:0.5)},rotate=90,anchor=center,font=\footnotesize},
    width=\linewidth,
	]%
    \addplot[red,mark=o] table[x=P,y=WER] {
    P      WER
    0.1    0.100301
    0.09   0.0384468
    0.08   0.0174551
    0.07   0.00972573
    0.06   0.00683013
    0.05   0.00608199
    0.04   0.00502942
    0.03   0.00350435
    0.02   0.00205727
    0.01   0.000355308
    };
    \addlegendentry{HP [[1922,50,16]]}

	\addplot[red,mark=*] table[x=P,y=WER] {
	P  WER
	0.14   0.483092
    0.12   0.0434405
    0.1    0.00112089
    0.08   7.75761e-05
    0.06   4.94918e-06
    0.04   9.55753e-08
	};
	\addlegendentry{HP, OSD [[1922,50,16]]}

	\addplot[black,mark=o] table[x=P,y=WER] {sims/res_ghpc33-882_24.txt};
    \addlegendentry{GHP [[882,24]]}
    
	\addplot[black,mark=*] table[x=P,y=WER] {sims/res_ghpc33-882_24-osd0.txt};
	\addlegendentry{GHP, OSD [[882,24]]}
	
	\end{loglogaxis}
	\end{tikzpicture}
	
	\caption{The WER performance of the $6$-limited $[[1922,50,16]]$ hypergraph product (HP) code and the $6$-limited $[[882,24]]$ generalized hypergraph product (GHP) code (\ref{code:hpc33-1922_50_16} and \ref{code:ghpc33-882_24} in Appendix~\ref{app:codes}). The HP code has an~error floor even under the BP-OSD-10 decoder.}
	\label{fg:ghp-vs-hp}
\end{figure}

Let us note that if the polynomial $b(x)$ is such that $g(x) = \gcd(b(x), x^\ell - 1)$ is an~irreducible factor of $x^\ell - 1$, then we can give a~more elegant formula for the dimension of the code $\cC(A,b)$ in some special cases. Indeed, in this case we have:
\begin{equation}\label{eq:GHP-dim}
    K = \deg g(x) (m+n-2\rk_F \varphi(A)), 
\end{equation}
where $F$ is the finite field $\F_2[x]/(g(x))$ and $\varphi(A)$ is the $F$-image of $A$ under the action of the map $\varphi\colon u(x) \mapsto u(x) \mod g(x)$. Since $\varphi(A)$ is a~matrix over $F$, it defines, as a~parity-check matrix, a~non-binary linear code over $F$ of dimension
\[
k_A = n - \rk_F \varphi(A).
\] 
At the same time, $b(x)$ defines the~cyclic code 
\[\cC_b = \cC_g = \{g(x)u(x) \mid u(x)\in \RC{2}{\ell} \}\]
of dimension $k_b = \deg g(x)$ as a~check polynomial. Thus formula~(\ref{eq:GHP-dim}) gives us the dimension $K$ of $\cC(A, b)$ in terms of the dimensions $k_A$ and $k_b$ for the two special cases shown below: 
\begin{enumerate}
    \item if $A$ is a square matrix, i.e. $m=n$, we have 
    \begin{equation}
      K=2k_A k_b;  
    \end{equation}
    \item if $A$ is a full rank matrix, i.e. $m = \rk_F \varphi(A)$, we have 
    \begin{equation}
      K=k_A k_b = (n - m) k_b.
    \end{equation}
\end{enumerate}

In the current work, we construct only codes that correspond to case~1 above. Such codes are defined by an~irreducible factor $b(x)$ of $x^\ell - 1$ and a~square matrix $A$ over $\RC{2}{\ell}$. To construct all the examples of QC GHP codes given in Appendix~\ref{app:codes} (codes~\ref{code:ghpc33-882_24}, \ref{code:ghpc35-882_48}, and \ref{code:ghpc33-1270_28}), we used the following procedure. First we fix a~low weight irreducible polynomial $b(x)\in\F_2[x]$ such that $b(x) \mid x^\ell -1$, and then randomly choose a~polynomial matrix $A$ subject to some constraints. Since we want to obtain sparse parity-check matrices $H_X$ and $H_Z$ for the code $\cC(A, b)$, we restrict each entry of $A$ to be \mbox{either} $0$ or a~monomial $x^i$, $0 \le i < \ell$. When the number of non-zero entries in each row and each column of $A$ is bounded above by $w$, this restriction guaranties that both $H_X$ and $H_Z$ are $(w+\deg b(x))$-limited matrices. Another restriction is related to the girth of the Tanner graphs $\cT_X$, $\cT_Z$ for $H_X$, $H_Z$. As you can see from Tab.~\ref{tab:codeparams} in Appendix~\ref{app:codes}, the girth of the Tanner graphs $\cT_X$, $\cT_Z$ for all the constructed QC GHP codes is equal to $6$, which means that $\cT_X$, $\cT_Z$ do not contain $4$-cycles. This restriction helps to improve the error-correcting performance under the BP decoder.     

\begin{figure}
	\begin{tikzpicture}
	\pgfplotsset{major grid style={dashed}} 
	\pgfplotsset{minor grid style={dotted}}
	\pgfplotsset{
        tick label style={font=\scriptsize},
        label style={font=\footnotesize},
        legend style={font=\scriptsize},
    }
	\begin{semilogyaxis}[
	title=Deep simulations of two GHP codes,
	tick label style={/pgf/number format/fixed},
	ymax=6*1e-1,
	ymin=1e-10,
    width=\linewidth,
    height=8cm,
	ylabel=WER,
	xlabel=Physical error rate,
    legend style={
    	cells={anchor=west},
    	legend pos= north west,
    },
    every axis y label/.style=
    {at={(ticklabel cs:0.5)},rotate=90,anchor=center,font=\footnotesize},
    grid=both,
	minor x tick num=4,
	minor y tick num=9,
	]%
	\addplot table[x=P,y=WER] {sims/res_ghpc-882_24-aug5-osd0-layered_nms-deep.txt};
	\addlegendentry{GHP [[882,24]]}
	\addplot[black,mark=*] table[x=P,y=WER] {sims/res_ghpc33-1270_28-aug5-osd0-layered_nms.txt};
	\addlegendentry{GHP [[1270,28]]}
	\addplot[red,mark=*] table[x=P,y=WER] {sims/surf25-mps.txt};
	\addlegendentry{Surf. MPS}
	\end{semilogyaxis}
	\end{tikzpicture}	

	\caption{A deep simulation on the depolarizing channel of the~\mbox{$6$-limited} [[882,24]] and [[1270,28]] QC GHP codes (\ref{code:ghpc33-882_24} and \ref{code:ghpc33-1270_28} in Appendix~\ref{app:codes}) under the BP-OSD-0 decoder. These codes do not have error floor down to $\mathrm{WER} = 10^{-10}$. The red curve is for the [[1201,1,25]] surface code on the~MPS-based decoder from~\cite{Bravyi:2014}.}
	\label{fg:deep}
\end{figure}

\begin{example}
In Fig.~\ref{fg:ghp-vs-hp} you can see the WER performance (under the BP-OSD decoder) of the two codes:  the $6$-limited [[1922,50,16]] hypergraph product (HP) code (\ref{code:hpc33-1922_50_16} in Appendix~\ref{app:codes}) and the $6$-limited [[882,24]] generalized hypergraph product (GHP) code (\ref{code:ghpc33-882_24} in Appendix~\ref{app:codes}). You can see from Fig.~\ref{fg:ghp-vs-hp} that the HP [[1922,50,16]] code has some error floor even under the BP-OSD decoder. We believe that this is due to a~large amount of low-weight non-degenerate codewords in this code.
\end{example}

In Fig~\ref{fg:deep} we can see the WER performance of the $6$-limited [[882,24]] and [[1270,28]] QC GHP codes (codes \ref{code:ghpc33-882_24} and \ref{code:ghpc33-1270_28} in Appendix~\ref{app:codes}) under the BP-OSD-0 decoder. We see that these codes do not have the error floor down to $\mathrm{WER} = 10^{-10}$. The red curve shows the corresponding WER performance of the $4$-limited [[1201,1,25]] surface code under the almost optimal MPS-based decoder from~\cite{Bravyi:2014}.



\section{Conclusion}

We proposed new OSD-like post-processing for the BP decoder that shows on some codes much better performance than all the modifications known to the authors. We also constructed several new generalized bicycle codes that show very good performance compared to the other known codes with similar parameters. We proposed a~new ansatz for quantum CSS codes and showed how to estimate the dimension of such codes in some special cases. Unfortunately, we have not found any nontrivial general lower bound on the minimum distance of such codes. We think that to find such a~bound is an~interesting open problem\footnote{Since the first version of this work appeared in arXiv, some lower bounds on the minimal distance have been obtained in~\cite{Hastings:2021,Panteleev&Kalachev:2020,Breuckmann:balanced:2021} for special cases of QC GHP codes $\cC(A, 1+x)$.} since this class contains some of the best known QLDPC codes, and their practical performance under the BP-OSD decoder is also quite good. Finally, we compared the performance of one of our codes from the new family and showed that it has better performance than a~relatively large surface code of similar code length even if this code is decoded by a~near-optimal decoder.

\begin{acknowledgments}
The authors would like to thank Dr.~Xuecang Zhang from Huawei Technologies for the support of this work and for useful discussions. This work was also supported by the Ministry of Science and Higher Education of the Russian Federation (Grant {\textnumero}~075-15-2020-801).

\end{acknowledgments}

\bibliography{quantum, coding}

\appendix

\section{Algebraic structure of the ring \texorpdfstring{$\RC{q}{\ell}$}{Rql}}\label{app:alg}
Let $\F_q$ be a finite field of characteristics 2. The algebraic structure of the ring $\RC{q}{\ell}$ is well studied in the coding literature (see, e.g.,~\cite{Ling2006}). Below we briefly review it.

First, let us consider the special case when $\ell$ is odd. In this case  the polynomial $x^\ell - 1$ factors into a~product of different irreducible polynomials over $\F_q$
\begin{equation}\label{eq:factors-co-prime}
x^\ell - 1  = f_1(x)\cdots f_s(x).
\end{equation}
This is true, since  
\[
\gcd\left((x^\ell - 1)', x^\ell - 1\right) = \gcd\left(\ell x^{\ell - 1}, x^\ell - 1\right) = 1,
\]
and the polynomial $x^\ell - 1$ is square-free. 

In the general case, we have $\ell = 2^e\ell'$, where $\ell'$ is odd. Hence it follows that 
\[x^\ell - 1 = x^{2^e\ell'} - 1 = (x^{\ell'} - 1)^{2^e}.\]
Moreover, since $\ell'$ is odd, we can apply factorization~(\ref{eq:factors-co-prime}) to the polynomial $x^{\ell'} - 1$ and obtain that 
\begin{equation}\label{eq:factors}
x^\ell - 1  = \bigl(f_1(x)\bigr)^{2^e}\cdots \bigl(f_s(x)\bigr)^{2^e}.
\end{equation}

Since the polynomials $(f_1(x)\bigr)^{2^e},\dots,(f_s(x)\bigr)^{2^e}$ are pairwise coprime, from the Chinese remainder theorem it follows that the ring $\RC{q}{\ell}$ is isomorphic to the direct product  
\begin{equation}\label{eq:isom-co-prime}
R_1\times \dots \times R_s
\end{equation}
of the rings $R_i = \F_q[x]/\bigl(f_i(x)\bigr)^{2^e}$, $i\in [s]$. 

When $\ell$ is odd we have $e=0$ and the rings $R_1,\dots,R_s$ are in fact fields, since the polynomials $f_1(x),\dots,f_s(x)$ are irreducible over $\F_q$.

\section{Matrices used for simulations}\label{app:codes}

All matrices for generalized hypergraph product and generalized bicycle codes that we used for simulations have the form $H_X=[A,B]$, $H_Z=[B^T,A^T]$ where $A$ and $B$ are quasi-cyclic matrices. Thus, to define the code, we specify matrices $A$ and $B$ in the polynomial form as matrices over $\RC{2}{\ell}$. 

For all codes presented here, we also provide lower and upper bounds on the minimum distance obtained either by methods similar to the ones from~\cite{Dumer:2017:dist} or by a~straightforward reduction of the minimum distance problem to a~mixed integer linear program and using the GNU Linear Programming Kit, Version~4.63, \mbox{\href{http://www.gnu.org/software/glpk/glpk.html}{http://www.gnu.org/software/glpk/glpk.html}}. 

\begin{enumerate}
    \renewcommand{\theenumi}{\Alph{enumi}}
    \renewcommand{\theenumii}{\arabic{enumii}}
    \renewcommand{\labelenumii}{\theenumi\theenumii)}
    \item {\bf Generalized bicycle (GB) codes.} The matrices $A$ and $B$ have form $A=(a(x))$, $B=(b(x))$, so here we specify the polynomials $a(x)$,  $b(x)$, and the circulant size $\ell$.
    \begin{enumerate}
        \item $[[254,28,d]]$ code ($\ell = 127$), $14 \le d \le 20$. \label{code:gbc-254_28}
\begin{align*}
a(x) & = 1+x^{15}+x^{20}+x^{28}+x^{66},\\
b(x) & = 1+x^{58}+x^{59}+x^{100}+x^{121}.
\end{align*}

        \item $[[126,28,8]]$ code ($\ell = 63$). \label{code:gbc-126_28}
\begin{align*}a(x)&= 1 + x + x^{14} + x^{16} + x^{22},\\
b(x) &= 1 + x^{3} + x^{13} + x^{20} + x^{42}.\end{align*}

        \item $[[48,6,8]]$ code ($\ell = 24$). \label{code:gbc-48_6_8}
\begin{align*}a(x)&= 1 + x^{2} + x^{8} + x^{15},\\
b(x) &= 1 + x^{2} + x^{12} + x^{17}.\end{align*}

        \item $[[46,2,9]]$ code ($\ell = 23$). \label{code:gbc-46_2_9}
\begin{align*}a(x)&= 1 + x^{5} + x^{8} + x^{12},\\
b(x) &= 1 + x + x^{5} + x^{7}.\end{align*}

        \item $[[180,10,d]]$ code ($\ell = 90$), $15 \le d \le 18$. \label{code:gbc44-180_10}
\begin{align*}a(x)&= 1 + x^{28} + x^{80} + x^{89},\\
b(x) &= 1 + x^{2} + x^{21} + x^{25}.\end{align*}

        \item $[[900,50,15]]$ code ($\ell = 450$). \label{code:gbc44-900_50}
\begin{align*}a(x)&= 1 + x^{97} + x^{372} + x^{425},\\
b(x) &= 1 + x^{50} + x^{265} + x^{390}.\end{align*}

    \end{enumerate}
    \item {\bf Generalized hypergraph product (GHP) codes.} Here the matrix $B$ is diagonal; $B=b(x)I_n$, where $I_n$ is the $n\times n$ identity matrix over the ring $\RC{2}{\ell}$.
    \begin{enumerate}
        \item $[[882,24,d]]$ code, $18 \le d \le 24$. The matrices $H_X$ and $H_Z$ are (3,6)-regular ($\ell = 63$).\label{code:ghpc33-882_24}
\begin{align*}A &= \left(\begin{smallmatrix}x^{27}&0&0&0&0&1&x^{54}\\
x^{54}&x^{27}&0&0&0&0&1\\
1&x^{54}&x^{27}&0&0&0&0\\
0&1&x^{54}&x^{27}&0&0&0\\
0&0&1&x^{54}&x^{27}&0&0\\
0&0&0&1&x^{54}&x^{27}&0\\
0&0&0&0&1&x^{54}&x^{27}
\end{smallmatrix}\right),\\
B &= (1 + x + x^{6})I_{7}.\end{align*}

        \item $[[882,48,16]]$ code. Half of the~columns for both $H_X$ and $H_Z$ matrices have weight $3$, another half have weight $5$. All the~rows have weight 8 ($\ell = 63$).\label{code:ghpc35-882_48}
\begin{align*}A &= \left(\begin{smallmatrix}x^{27}&0&0&1&x^{18}&x^{27}&1\\
1&x^{27}&0&0&1&x^{18}&x^{27}\\
x^{27}&1&x^{27}&0&0&1&x^{18}\\
x^{18}&x^{27}&1&x^{27}&0&0&1\\
1&x^{18}&x^{27}&1&x^{27}&0&0\\
0&1&x^{18}&x^{27}&1&x^{27}&0\\
0&0&1&x^{18}&x^{27}&1&x^{27}
\end{smallmatrix}\right),\\
B &= (1 + x + x^{6})I_{7}.\end{align*}

        \item $[[1270,28,d]]$ code, $16 \le d\le 46$. The matrices $H_X$ and $H_Z$ are (3,6)-regular ($\ell = 127$). \label{code:ghpc33-1270_28}
\begin{align*}A &= \left(\begin{smallmatrix}1&0&x^{51}&x^{52}&0\\
0&1&0&x^{111}&x^{20}\\
1&0&x^{98}&0&x^{122}\\
1&x^{80}&0&x^{119}&0\\
0&1&x^{5}&0&x^{106}
\end{smallmatrix}\right),\\
B &= (1 + x + x^{7})I_{5}.\end{align*}

    \end{enumerate}
    \item {\bf Hypergraph product (HP) codes.} Each hypergraph product code in our simulations is constructed from a single cyclic code defined by its parity polynomial $h(x)$ and the length $\ell$.
    \begin{enumerate}
        \item $[[7938,578,16]]$ code. The matrices $H_X$ and $H_Z$ are (5,10)-regular, and we have: \label{code:hpc55-7938_578_16}
$$\ell = 63, \quad h(x)=1 + x^{3} + x^{34} + x^{41} + x^{57}.$$

        \item $[[1922,50,16]]$ code. The matrices $H_X$ and $H_Z$ are (3,6)-regular, and we have: \label{code:hpc33-1922_50_16}
$$\ell=31,\quad h(x)=1 + x^{2} + x^{5}.$$

    \end{enumerate}
    \item {\bf Haah's cubic codes.} We used the $[[1024,30,d]]$ Haah's cubic code on the $8\times 8\times 8$ lattice from~\cite[Code~1]{Haah:2011}, $13 \le d \le 32$. \label{code:haah8}
    \item {\bf Hyperbicycle (HB) codes.} We used the $[[900,50,14]]$ hyperbicycle code from~\cite[Example~8]{Kovalev&Pryadko:2012}, \cite[Example~6]{Kovalev&Pryadko:HBP:2013}.\label{code:hbc900_50_14}
    \item {\bf Homological product (HMP) codes.} We used one of the randomly constructed $[[49,1,9]]$ homological product codes from~\cite{Bravyi:HMP:2014}.\label{code:hmp49_1_9}
\end{enumerate}

\begin{table}[ht]
    \centering
        \caption{Codes parameters}
        
\scalebox{0.9}{
\begin{tabular}{|c|c|c|c|c|c|c|c|}
\hline
 Code & $N$ & $K$ & $d$ & rate & $w_r$ & $w_c$ &  girth \\
\hline
 \ref{code:gbc-254_28} & 254 & 28 & 14--20 & 0.110 & 10 & 5 &  6 \\
\hline
 \ref{code:gbc-126_28} & 126 & 28 & 8 & 0.222 & 10 & 5 & 4 \\
\hline
 \ref{code:gbc-48_6_8} & 48 & 6 & 8 & 0.125 & 8 & 4 & 4 \\
\hline
 \ref{code:gbc-46_2_9} & 46 & 2 & 9 & 0.043 & 8 & 4 & 4 \\
\hline
 \ref{code:gbc44-180_10} & 180 & 10 & 15--18 & 0.056 & 8 & 4 & 6 \\
\hline
 \ref{code:gbc44-900_50} & 900 & 50 & 15 & 0.056 & 8 & 4 & 6 \\
\hline
 \ref{code:ghpc33-882_24} & 882 & 24 & 18--24 & 0.027 & 6 & 3 & 6 \\
\hline
 \ref{code:ghpc35-882_48} & 882 & 48 & 16 & 0.054 & 8 & 3,5 & 6 \\
\hline
 \ref{code:ghpc33-1270_28} & 1270 & 28 & 16--46 & 0.022 & 6 & 3 & 6 \\
\hline

 \ref{code:hpc55-7938_578_16} & 7938 & 578 & 16 & 0.073 & 10 & 5 & 6 \\
\hline
 \ref{code:hpc33-1922_50_16} & 1922 & 50 & 16 & 0.026 & 6 & 3 & 6 \\
\hline
 \ref{code:haah8} & 1024 & 30 & 13--32 & 0.029 & 8 & 4 & 4 \\
\hline
 \ref{code:hbc900_50_14} & 900 & 50 & 14 & 0.056 & 8 & 4 & 4 \\
\hline
 \ref{code:hmp49_1_9} & 49 & 1 & 9 & 0.020 & 6,8 & 6,8 & 4 \\
\hline
\end{tabular}    
}
    \label{tab:codeparams}
\end{table}

\section{Additional Simulations}\label{app:sims}

In this appendix, we show some additional simulation results of the BP-OSD decoder on the depolarizing channel. 

In~Fig.~\ref{fg:thershold}, we demonstrate the WER performance of several $6$-limited generalized bicycle (GB) codes with the parameters $[[2^{s+1} - 2, 2s]]$, $s\in\N$, under the BP-OSD-0 decoder. As~one can see from the curves, these codes have the~threshold  $p_{\mathrm{GB}} \approx 15\%$, which is quite close to the corresponding threshold $p_{\mathrm{S}} \approx 18\%$ of the surface codes from~\cite{Bravyi:2014}. The codes from this family were constructed in the same way as the GB codes from Subsection~\ref{sc:GB-constr}, where for each $s\in\N$ the corresponding syndrome code $\cC_g$ is the cyclic Hamming $[2^s - 1, s, 3]$ code defined by some irreducible polynomial $g(x)\in \F_2[x]$ of~degree $s$.

In all our previous examples, we considered only CSS codes. In fact, as it is shown in Subsection~\ref{sc:BP-OSD}, the BP-OSD decoder can be applied to any stabilizer code. In~Fig.~\ref{fg:hbc-non-css}, we demonstrate the WER performance of the BP-OSD-0 decoder on the $5$-limited cyclic non-CSS [[126,2,12]] code. This code is defined by the parity-check matrix $H=[H_X\mid H_Z]$, where $H_X$ and $H_Z$ are the $126\times 126$ circulant matrices represented respectively by the polynomials: $1+x^{71}+x^{55}$ and $1+x^{40}+x^{86}$. As~we can see, the WER gain of OSD-0 post-processing in this case is at least three orders of magnitude.

\begin{figure}
	\begin{tikzpicture}
	\pgfplotsset{major grid style={dashed}} 
	\pgfplotsset{minor grid style={dotted}}
	\pgfplotsset{
        tick label style={font=\footnotesize},
        label style={font=\footnotesize},
        legend style={font=\scriptsize},
    }
	\begin{semilogyaxis}[
	title=Threshold for GB codes (BP-OSD-0),
	ylabel=WER,
	xlabel=Physical error rate,
	ymax=1,
	ymin=1e-1,
	legend style={
		cells={anchor=west},
		legend pos=north west,
	},
    every axis y label/.style=
    {at={([xshift=5pt]ticklabel cs:0.5)},rotate=90,anchor=center,font=\footnotesize},	
	grid=both,
	minor x tick num=4,
	width=\linewidth,
	]%
	\addplot table[x=P,y=WER] {sims/res_gbc33ham-126_12_10-osd.txt};
	\addlegendentry{[[126,12]]}
	\addplot table[x=P,y=WER] {sims/res_gbc33ham-254_14_16-osd.txt};
	\addlegendentry{[[254,14]]}
	\addplot table[x=P,y=WER] {sims/res_gbc33ham-510_16_17-osd.txt};
	\addlegendentry{[[510,16]]}
	\addplot table[x=P,y=WER] {sims/res_gbc33ham-1022_18_17-osd.txt};
	\addlegendentry{[[1022,18]]}
	\addplot table[x=P,y=WER] {sims/res_gbc33ham-8090_24_17-osd.txt};
	\addlegendentry{[[8190,24]]}

	\end{semilogyaxis}
	\end{tikzpicture}
	\caption{The WER performance on the depolarizing channel for five $6$-limited GB codes under the \mbox{BP-OSD-0} decoder. The threshold is $p_{\mathrm{GB}} \approx 15\%$. The corresponding threshold of the $4$-limited surface codes~\cite{Bravyi:2014} is $p_{\mathrm{S}} \approx 18\%$.	}
	\label{fg:thershold}
\end{figure}

\begin{figure}
	\begin{tikzpicture}
	\pgfplotsset{major grid style={dashed}} 
	\pgfplotsset{minor grid style={dotted}}
	\pgfplotsset{
        tick label style={font=\footnotesize},
        label style={font=\footnotesize},
        legend style={font=\scriptsize},
    }
	\begin{loglogaxis}[
	ylabel=WER,
	xlabel=Physical error rate,
	legend style={
		cells={anchor=west},
		legend pos=north west,
	},
	grid=both,
	title={Cyclic non-CSS code (BP-OSD-0)},
	xtick={1e0, 1e-1, 1e-2},
	extra x ticks={5e-2,2e-1},
	enlarge y limits=.01,
	every axis y label/.style=
    {at={([xshift=-2pt]ticklabel cs:0.5)},rotate=90,anchor=center,font=\footnotesize},
	]%

    \addplot table[x=P,y=WER] {sims/res_gbc5-126_4.txt};
    \addlegendentry{BP}
    \addplot table[x=P,y=WER] {sims/res_gbc5-126_4-osd0.txt};
    \addlegendentry{BP-OSD-0}
	\end{loglogaxis}
	\end{tikzpicture}
	\caption{The WER performance of the BP-OSD-0 decoder on the $5$-limited cyclic non-CSS [[126,2,12]] code defined by the parity-check matrix $H=[H_X\mid H_Z]$, where $H_X$ and $H_Z$ are the $126\times 126$ circulant matrices represented respectively by the polynomials: $1+x^{71}+x^{55}$ and $1+x^{40}+x^{86}$.}
	\label{fg:hbc-non-css}
\end{figure}

\end{document}